\documentclass[reprint,amsmath,amssymb,aps,prl]{revtex4-2}
\usepackage{times}
\usepackage{physics,tikz,bm}
\usepackage[normalem]{ulem}
\usepackage[pdftex,colorlinks,linkcolor={blue!75!black},citecolor={blue!75!black},urlcolor={blue!75!black}]{hyperref}

\begin{document}

\title{Ultra spectral sensitivity and non-local bi-impurity bound states from quasi-long-range non-hermitian skin modes}
\author{Chang Shu}
\affiliation{Department of Physics, University of Michigan, Ann Arbor, Michigan 48109, United States}
\author{Kai Zhang}
\email{phykai@umich.edu}
\affiliation{Department of Physics, University of Michigan, Ann Arbor, Michigan 48109, United States}
\author{Kai Sun}
\email{sunkai@umich.edu}
\affiliation{Department of Physics, University of Michigan, Ann Arbor, Michigan 48109, United States}

\date{\today}

\begin{abstract}
A fundamental tenet of quantum mechanics is that the energy spectrum of a quantum system shall remain stable against infinitesimally weak and spatially confined perturbations. This stability forms the foundation of key concepts such as perturbation theory and Feynman's diagrammatic expansions. In this article, we demonstrate that this principle of spectral stability fails in non-Hermitian systems at the thermodynamic limit. Consider, for instance, a non-interacting non-Hermitian system $\hat{H}_0$ with a couple of point-like impurities, each of which introduces a local short-range potential $\hat{V}_i$ with $i=1, \ldots, n$ labeling the impurities. If the impurity potentials are sufficiently weak, introducing a single impurity will not alter the spectrum; that is, $\hat{H}_0$ and $\hat{H}_0 + \hat{V}_1$ have nearly identical energy spectra, with deviations proportional to $\hat{V}_1$ and negligible for small $\hat{V}_1$. However, if a second impurity is introduced, $\hat{H}_0 + \hat{V}_1 + \hat{V}_2$, we find that no matter how weak these local potentials are, as long as the distance between them is sufficiently large, significant alterations in the energy spectrum can arise, directly contradicting the traditional expectation of a stable spectrum. Remarkably, this phenomenon is a non-local effect, and the impact of the perturbations increases exponentially with the distance between the two impurities. In other words, although the Hamiltonian is entirely local, its energy spectrum, which is blind to the presence of a single infinitesimally weak impurity, is capable of detecting the presence of two infinitesimally weak impurities separated by a large distance in space. Using Green's function techniques, we uncover the origin of this spectral sensitivity, which arises from the formation of non-local bi-impurity bound states: non-local stationary states with wavepackets propagating back-and-forth between the two impurities. We provide an analytic theory to identify and characterize such spectral instabilities, showing perfect agreement with numerical solutions. Furthermore, through time-dependent dynamical simulations, we show that these non-local bi-impurity bound states could be directly observed in experiments by tracking the time evolution of wavepackets.
\end{abstract}

\maketitle

\section{Introduction}

The study of non-Hermitian Hamiltonians has paved the way for the discovery of novel phenomena and new principles of physics across diverse open classical and quantum systems, covering photonic crystals, cold-atom arrays, and mechanical/acoustic metamaterials~\cite{Ashida2020}. Among the recent breakthroughs, the concept of sensitivity, where the system exhibits a strong response to certain minute perturbations, is a crucial property of non-Hermiticity and has garnered significant interest in optical and quantum devices~\cite{Ganainy2018}. One pronounced example is that the system's response follows a square-root function of external controlling parameters when the system is near exceptional points—a type of non-Hermitian degeneracy—making it useful in sensing applications~\cite{Hodaei2017,YangLan2019}. Another pivotal exotic sensitivity is related to the non-Hermitian skin effect (NHSE)~\cite{Yao2018,Kunst2018PRL,GongPRX2018,Slager2020PRL,Kai2020,Okuma2020PRL,ZSaGBZPRL,KawabataPRB2020,XuePeng2020,Thomale2020,Ghatak2020,LiLH2020NC,LuMing2021,Murakami2022,HuiJiang2022,Kai2023PRB,HYWang2024,Fulga2024NP,OkumaSatoReview,LeeCHReview,YFChen2022Review}, a phenomenon where the majority of eigenstates are localized at the system's boundaries. In one-dimensional (1D) systems, the eigenspectrum and wavefunctions are highly sensitive to changes in boundary conditions. Even minor boundary links can dramatically alter the entire spectrum, which has been utilized for quantum sensors~\cite{Budich2020PRL,budich2023,Fulga2024NP}. Furthermore, when the NHSE is symmetry protected~\cite{Okuma2020PRL}, the spectrum exhibits infinitesimal instability to spatially uniform perturbations that break the symmetry~\cite{Sato2019PRL}. These sensitivities are hallmarks of critical phenomena in various 1D NHSE systems~\cite{LiLH2020NC}, which can be well understood within the 1D generalized Bloch framework~\cite{ZSaGBZPRL,LiLH2020NC,Yao2018,Murakami2019PRL,KawabataPRB2020}.
At these critical phases, the system exhibits scale-free localization behavior~\cite{LiLH2020NC,LLHScaleFree2021,Kawabata2023PRX,ZhangYi2023PRB}, where the localization length of the wavefunction scales with the system size. However, these sensitivities typically require 1D systems to be at or near specific points in the phase diagram, such as the non-reciprocal point or exceptional points, with perturbations arranged in specific configurations, such as finely tuned boundary links or symmetry-breaking couplings. These requirements make such effects less convenient to be observed in experiments and to be utilized in applications.

Recent studies in non-Bloch band theory have uncovered that, without any need for fine-tuning, non-Hermitian skin modes in two and higher dimensions can exhibit quasi-long-distance algebraic (power-law) decay~\cite{KaiASE2024}, contrasting the conventional short-range exponential decay observed in typical 1D NHSE~\cite{Yao2018}. Due to the non-local nature of power-law functions, this skin-mode paradigm prompts an essential inquiry: could there be non-local sensitivity associated with such algebraic modes? Additionally, in two and higher dimensions, the open boundary spectrum becomes highly sensitive to geometric variables, particularly the shape and spatial aspect ratio of the open boundary geometry~\cite{Kai2022NC,DingKun2023PRL,QYZhou2023NC,HaiPing2024arXiv}. This raises critical questions: What is the role of the spatial ratio in such critical behavior? Can we develop an analytic formula to elucidate it? Given that the experimental realization of algebraic skin effects is relatively straightforward, requiring only some on-site dissipation, the experimental impacts and signatures of such quasi-long-range modes and non-local phenomena naturally emerge as significant open questions.

In this study, we focus on the question of whether infinitesimally weak and spatially confined perturbations can significantly alter the energy spectrum. To ensure generality, we avoid special points in the phase space that require fine-tuning to reach. In a Hermitian system, the answer to this question is negative. For example, consider a free particle moving in a $d$-dimensional space, and introduce a spatially confined (local) potential trap as a perturbation. To have a significant impact on the energy spectrum, the potential trap must exceed a certain threshold, enabling the particle to be trapped by the potential to form local bound states. For very weak potential traps below such thresholds, their impact on the energy spectrum is negligible. For non-Hermitian systems, however, the quasi-long-range nature of the skin modes provides an alternative pathway to significantly alter the energy spectrum, utilizing the diverging non-local response induced by such NHSE. We can simply introduce two (or more) point-like impurities, $\hat{V}_1$ and $\hat{V}_2$, both local and weak. Instead of enhancing the strength of these impurity potentials, we use the distance between the two impurities as our control parameter. As illustrated in Fig.~\ref{fig:1}, for a wide range of non-Hermitian systems, no matter how weak the impurity potentials are, as long as the distance between them is sufficiently large, infinitely many new eigenstates are generated by these small perturbations, with eigenenergies far from the spectrum of the unperturbed Hamiltonian. In other words, the energy spectrum is significantly changed.

We further proved that the origin of this spectral sensitivity comes from the diverging non-local response characterized by the Green's function $G_0(E;\mathbf{r}_1,\mathbf{r}_2)G_0(E;\mathbf{r}_2,\mathbf{r}_1)$, where $G_0$ is the single-particle Green's function of the unperturbed Hamiltonian at energy $E$. $\mathbf{r}_1$ and $\mathbf{r}_2$ are the locations of the two impurities, respectively. We prove that, in general, the impact of introducing two weak and local impurity potentials is measured by $V \sqrt{|G_0(E;\mathbf{r}_1,\mathbf{r}_2)G_0(E;\mathbf{r}_2,\mathbf{r}_1)|}$, where $V$ is the strength of the impurity potentials. In Hermitian systems, as well as typical 1D non-Hermitian systems with non-reciprocal skin effects, $G_0(E;\mathbf{r}_1,\mathbf{r}_2)G_0(E;\mathbf{r}_2,\mathbf{r}_1)$ exhibits no divergence as we vary the distance between the impurities, as long as $E$ remain off-resonance, and thus the impact of the impurities vanishes for small $V$. In contrast, for a wide range of non-Hermitian systems, e.g., 2D systems with algebraic skin effects, $G_0(E;\mathbf{r}_1,\mathbf{r}_2)G_0(E;\mathbf{r}_2,\mathbf{r}_1)$ diverges exponentially as the distance between the two impurities increases. In such scenarios, no matter how weak the impurity potential is, as long as they are sufficiently apart in space, the impact of $V$ becomes non-perturbative and leads to a dramatic impact on the energy spectrum. These dynamical characteristics of non-local bound states can be directly probed in experiments and serve as experimental signatures for their detection.

Using this Green's function formula, we identify the phase diagram of this spectrum sensitivity phenomenon, which is dictated by two factors: the geometry aspect ratio of the non-Hermitian system and the relative locations of the two impurities. In two dimensions, different aspect ratios correspond to distinct thermodynamic limits, resulting in distinct spectral sensitivities, and our analytic solution shows perfect agreement with numerical solutions. Additionally, we further propose a dynamical consequence of this bi-impurity-induced sensitivity: the two impurities create a long-lived communication channel for wavepackets to propagate back-and-forth between them, forming non-local and dynamically bound states, which are the origin of the new eigenstates induced by the impurities.

\section{A model example}
\begin{figure}[t]
    \centering
    \includegraphics[width=\linewidth]{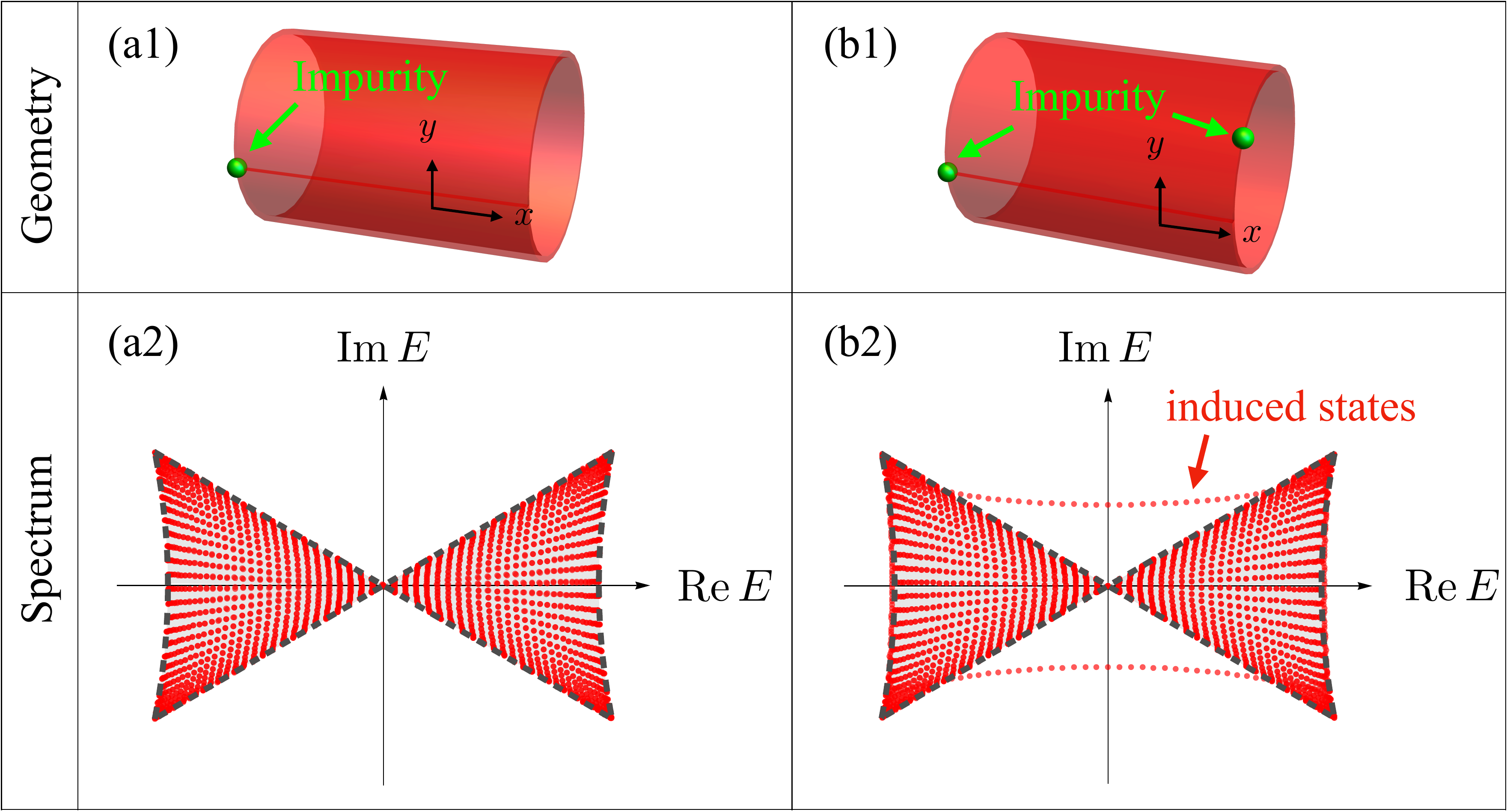}
    \caption{\textbf{Response of the energy spectrum to extremely weak local impurities.}
    Here we examine a non-Hermitian lattice model defined in Eq.~\eqref{eq:2DTBModel}. For simplicity, we consider the cylinder geometry and introduce impurities near the edges with extremely weak on-site impurity potentials. (a1) A single weak impurity potential with strength $V_1 = \frac{1}{100}$ is placed at the left edge of the cylinder. (a2) The red dots mark the numerically calculated energy spectrum in the presence of this impurity, and the dashed black lines indicate the boundary of the energy spectrum without any impurity. The perfect agreement between the spectrum with (red dots) and without (black dashed lines) the impurity indicates that a single impurity acts as a small perturbation to the energy spectrum.
    (b1) Illustration of introducing two impurities at the opposite edges of the cylinder with impurity strengths $V_1 = V_2 = \frac{1}{100}$.
    (b2) In the presence of the two impurities, two lines of new eigenstates are induced, marked as ``induced states" in the figure, with energies away from the unperturbed energy spectrum. The emergence of these new eigenstates indicates that, despite their small values, the impact of such local impurity potentials is non-perturbative and can significantly alter the energy spectrum. The numerical calculation is performed using the model on a cylinder with system size $L_x \times L_y = 60 \times 60$. The Hamiltonian parameters are set to $t_x = i$, $t_y = 0$, $t_{xy} = 3$, and $u = -2i$.}
    \label{fig:1}
\end{figure}

We begin the discussion by demonstrating such infinitesimal spectrum sensitivity in a simple model system. Considering the following reciprocal non-Hermitian tight-binding model on a cylinder with size $L_x\times L_y$:
\begin{equation}\label{eq:2DTBModel}
    \begin{aligned}
        \hat{H}_{0} & = \sum_{x,y} t_x \, (\hat{c}^{\dagger}_{x+1,y} \hat{c}_{x,y} + \text{H.c.}) + t_y \, (\hat{c}^{\dagger}_{x,y+1} \hat{c}_{x,y} + \text{H.c.}) \\
        & + t_{xy} \, (\hat{c}^{\dagger}_{x+1,y+1} \hat{c}_{x,y} + \text{H.c.}) + u \, \hat{c}^{\dagger}_{x,y} \hat{c}_{x,y}.
    \end{aligned}
\end{equation}
Here, we adopt open boundary condition (OBC) along the $x$ and periodic boundary condition (PBC) along the $y$ direction. Here, $t_x$,$t_y$, and $t_{xy}$ represent coupling strengths along the $x$, $y$, and $x+y$ directions, respectively, and $u$ denotes the onsite potential strength. These parameters are generally complex-valued and represent dissipative couplings and onsite gain/loss. Such reciprocal dissipative non-Hermitian Hamiltonians can be realized by mechanical/phononic metamaterials~\cite{DingKun2023PRL,QYZhou2023NC} or ultracold Fermi gas~\cite{GBJo2023arXiv}. 
As illustrated in Fig.~\ref{fig:1}(a1), we first place a weak impurity on the left edge of the cylinder, represented by $\hat{V}=V_1\delta_{\mathbf{r},\mathbf{r}_1}$. 
The resulting spectrum [indicated by the red dots in Fig.~\ref{fig:1} (a2)] is almost identical to the original impurity-free spectrum [represented by the continuum region bounded by the black dashed lines]. 
This result indicates that a single impurity can be treated as a perturbation, with no sensitive response in the spectrum.

In sharp contrast, when we introduce a second weak impurity on the opposite edge of the cylinder, as illustrated in Fig.~\ref{fig:1}(b1), the spectrum undergoes an abrupt change, with two induced spectrum lines dramatically shifted from the cylindrical spectrum, as shown in Fig.~\ref{fig:1}(b2). 
We dub the setting of a cylinder with two generic, distant impurities the bi-impurity cylinder. 
One example is placing two onsite potentials at two edges, as illustrated in Fig.~\ref{fig:1}(b1), represented as 
\begin{equation}\label{eq:bi_imp}
\hat{V}=V_1\delta_{\mathbf{r},\mathbf{r}_1}+V_2\delta_{\mathbf{r},\mathbf{r}_2}
\end{equation}
Remarkably, the infinitesimal instability of cylinder spectrum to the two edge impurities reveals a novel type of spectral sensitivity that is fundamentally different from previously known types~\cite{LiLH2020NC,ZSaGBZPRL,Okuma2020PRL,Kawabata2023PRX}. 
It is worthwhile to emphasize that to observe such infinitesimal sensitivity, the location of the two impurities doesn't need to be at the edge of the cylinder. Two distant bulk sites can produce the same effect. However, because the impact of the bi-impurity increases with the distance between them, for a fixed system size, using edge sites maximizes its spectrum sensitivity.  

For a two-dimensional system, its thermodynamic limit is not uniquely defined but depends on its aspect ratio. Here we define aspect ratio $\alpha$ as $L \equiv L_x = \alpha L_y$ and set the thermodynamic limit to $L\to \infty$. For Hermitian systems, the value of $\alpha$ becomes irrelevant at $L\to \infty$. However, for non-Hermitian systems, each value of $\alpha$ defines a different and unique thermodynamic limit.  In addition, for a given thermodynamic limit with fixed $\alpha$, the ratio of the distances between the two impurities in the $x$ and $y$ directions, defined as $g=(y_1-y_2)/L$, acts as another key degree of freedom in bi-impurity configuration given by Eq.~\eqref{eq:bi_imp}. 
We find that the spectral sensitivity of the cylinder spectrum strongly depends on these two geometric features emergent in two and higher dimensions (See Supplementary Note I). 
These observations lead us to the following crucial questions: What is the underlying reason behind this bi-impurity-induced sensitivity? Is there a formula to quantitatively identify the roles of $\alpha$ and $g$ under different thermodynamic limits? 

\section{the physics origin of the sensitivity: the divergence of the forth-back propagator}
This infinitesimal spectral sensitivity originates from the divergence of a non-local response, characterized by the forth-back propagator. To demonstrate this, we prove two mathematical theorems that apply generically to any Hermitian or non-Hermitian systems with translation symmetry: (1) For a single, weak, and local impurity at the location $\mathbf{r}$, its impact on the spectrum is measured by $V G_0(E; \mathbf{r},\mathbf{r})$, where $V$ is the strength of the impurity potential and $G_0$ is the single-particle Green's function of the unperturbed Hamiltonian at energy $E$. If $E$ is not in resonance with any eigenstates of the unperturbed Hamiltonian, the autocorrelation function $G_0(E; \mathbf{r},\mathbf{r})$ is generally finite. Thus, if $V$ is sufficiently weak, its impact must be vanishingly small.
(2) If two impurities with weak and local potentials are present at $\mathbf{r}_1$ and $\mathbf{r}_2$, the impact of the bi-impurity perturbation is controlled by $V \sqrt{|G_0(E; \mathbf{r}_1,\mathbf{r}_2) G_0(E, \mathbf{r}_2,\mathbf{r}_1)|}$. Besides the strength of the impurity potential, the impact is also controlled by the non-local response function $G_0(E; \mathbf{r}_1,\mathbf{r}_2) G_0(E, \mathbf{r}_2,\mathbf{r}_1)$, which we term the forth-back propagator. For Hermitian systems, as well as typical 1D non-Hermitian systems with non-reciprocal skin effects, this forth-back propagator converges as long as $E$ is off-resonance. Thus, the impact of weak impurity potentials remains perturbative.

However, in the model described in Eq.~\eqref{eq:2DTBModel} and many other non-Hermitian systems, especially those that exhibit algebraic skin effects, the forth-back propagator diverges exponentially as the distance between the two impurities increases. Consequently, no matter how weak the impurity potentials are, their impact becomes non-perturbative if the distance between them is sufficiently large. This divergence of the forth-back propagator represents a diverging non-local response, which is the origin of the spectral instability against two infinitesimally weak local impurities. Such a diverging non-local response is a non-Hermitian effect and it arises in systems where the Hamiltonian is entirely local. It is the non-Hermitian eigenstates of these systems that enable information to propagate over long distances between the two impurities, thus facilitating the diverging non-local response.

To prove the theorem, we consider the correlation function for the full Hamiltonian $\hat{H} = \hat{H}_0 + \hat{V}$, which can be expressed as
\begin{align}
\hat{G} = (E \hat{I} - \hat{H})^{-1} = (\hat{I} - \hat{G}_0 \hat{V})^{-1} \hat{G}_0
\label{eq:green_function}
\end{align}
where $\hat{G}_0 = (E - \hat{H}_0)^{-1}$ is the Green's function of the unperturbed Hamiltonian $\hat{H}_0$.
Because we are focusing on whether $\hat{H}$ supports new eigenstates with energies away from the spectrum of $\hat{H}_0$, we just need to focus on the values of $E$ beyond the eigenvalues of $\hat{H}_0$, i.e., $E$ is off-resonance for $\hat{H}_0$. For such off-resonance energies, $\hat{G}_0$ is in general non-singular.

From Eq.~\eqref{eq:green_function}, it is straightforward to realize that the difference between $\hat{G}$ and $\hat{G}_0$ is measured by the operator $(\hat{I} - \hat{G}_0 \hat{V})^{-1}$. Thus, if all the eigenvalues of $\hat{G}_0 \hat{V}$ are significantly smaller than $1$, $\hat{G} \approx \hat{G}_0$, and the spectrum of $\hat{H}$ and $\hat{H}_0$ would coincide. In contrast, if $\hat{G}_0 \hat{V}$ contains large eigenvalues of the order of $1$ or larger, $\hat{G}$ and $\hat{G}_0$ will exhibit significant deviations beyond the perturbative regime. To rigorously quantify this conclusion, we define the spectral radius ($\rho$) of an operator as the maximum of the absolute values of its eigenvalues, and the condition of spectrum stability is given by $\rho(\hat{G}_0 \hat{V}) \ll 1$.

As shown in the Methods section, for a single impurity at location $\mathbf{r}_1$, $\hat{V} = V_1 \delta_{\mathbf{r}, \mathbf{r}_1}$, the operator $\hat{G}_0 \hat{V}$ has only one nonzero eigenvalue $V_1 G_0(E; \mathbf{r}_1,\mathbf{r}_1)$, whose value vanishes at the limit $V\to 0$. If two impurities are present at locations $\mathbf{r}_1$ and $\mathbf{r}_2$, with $\hat{V} = V_1 \delta_{\mathbf{r}, \mathbf{r}_1} + V_2 \delta_{\mathbf{r}, \mathbf{r}_2}$, $\hat{G}_0 \hat{V}$ has two non-zero eigenvalues, which are the eigenvalues of the following $2 \times 2$ matrix:
\begin{align}
\begin{pmatrix}
    V_1 G_0(E; \mathbf{r}_1, \mathbf{r}_1) & V_2 G_0(E; \mathbf{r}_1, \mathbf{r}_2) \\
    V_1 G_0(E; \mathbf{r}_2, \mathbf{r}_1) & V_2 G_0(E; \mathbf{r}_2, \mathbf{r}_2)
\end{pmatrix}
\label{eq:biimp_response}
\end{align}
If $V_1 = V_2 = V$ and $G_0(E; \mathbf{r}_1, \mathbf{r}_1) = G_0(E; \mathbf{r}_2, \mathbf{r}_2) = G_0(E, 0)$, the eigenvalues are
\begin{align}
\xi_{\pm} = V \left( G_0(E, 0) \pm \sqrt{G_0(E; \mathbf{r}_1, \mathbf{r}_2) G_0(E; \mathbf{r}_2, \mathbf{r}_1)} \right).
\end{align}
In addition to the single impurity contribution $G_0(E, 0)$, the eigenvalues also depend on the square root of the forth-back propagator, $G_0(E; \mathbf{r}_1, \mathbf{r}_2) G_0(E; \mathbf{r}_2, \mathbf{r}_1)$, which describes the propagation of wavepackets between the two impurities.

As discussed above, for off-resonance $E$, $G_0(E, 0)$ is generally finite. As for the forth-back propagator, in Hermitian systems or non-Hermitian systems without skin modes, $G_0(E; \mathbf{r}_1, \mathbf{r}_2)$ is typically finite. Thus, for weak enough impurity potentials, $\rho(\hat{G}_0 \hat{V}) \ll 1$ and the spectrum remains stable. For non-Hermitian systems with NHSE, if all skin modes are localized at the same end of the system, $G_0(E; \mathbf{r}_1, \mathbf{r}_2)$ increases exponentially as $\mathbf{r}_1 - \mathbf{r}_2 \to +\infty$, while $G_0(E; \mathbf{r}_2, \mathbf{r}_1)$ decreases exponentially to zero as $\mathbf{r}_2 - \mathbf{r}_1 \to -\infty$. Thus, these two exponential functions cancel each other, and the forth-back propagator remains finite, maintaining spectral stability.

However, if the NHSE has skin modes localized on both edges/ends of the system, $G_0(E; \mathbf{r}_1, \mathbf{r}_2)$ and $G_0(E; \mathbf{r}_2, \mathbf{r}_1)$ will both diverge exponentially to infinity as the distance between $\mathbf{r}_1$ and $\mathbf{r}_2$ increases. In this scenario, the forth-back propagator diverges, and thus, no matter how weak the impurity potentials are, their impact on the Green's function (and thus the energy spectrum) will be non-perturbative, as long as the distance between the two impurities is sufficiently large, such that: 
\begin{align}
    V_* &\sim \frac{1}{\sqrt{|G_0(E; \mathbf{r}_1, \mathbf{r}_2) G_0(E; \mathbf{r}_2, \mathbf{r}_1)|}}.
\label{eq:senscriterion}
\end{align}
This is the physical origin of the infinitesimal spectral instability induced by the non-local response.

\begin{figure}[t]
    \centering
    \includegraphics[width=\linewidth]{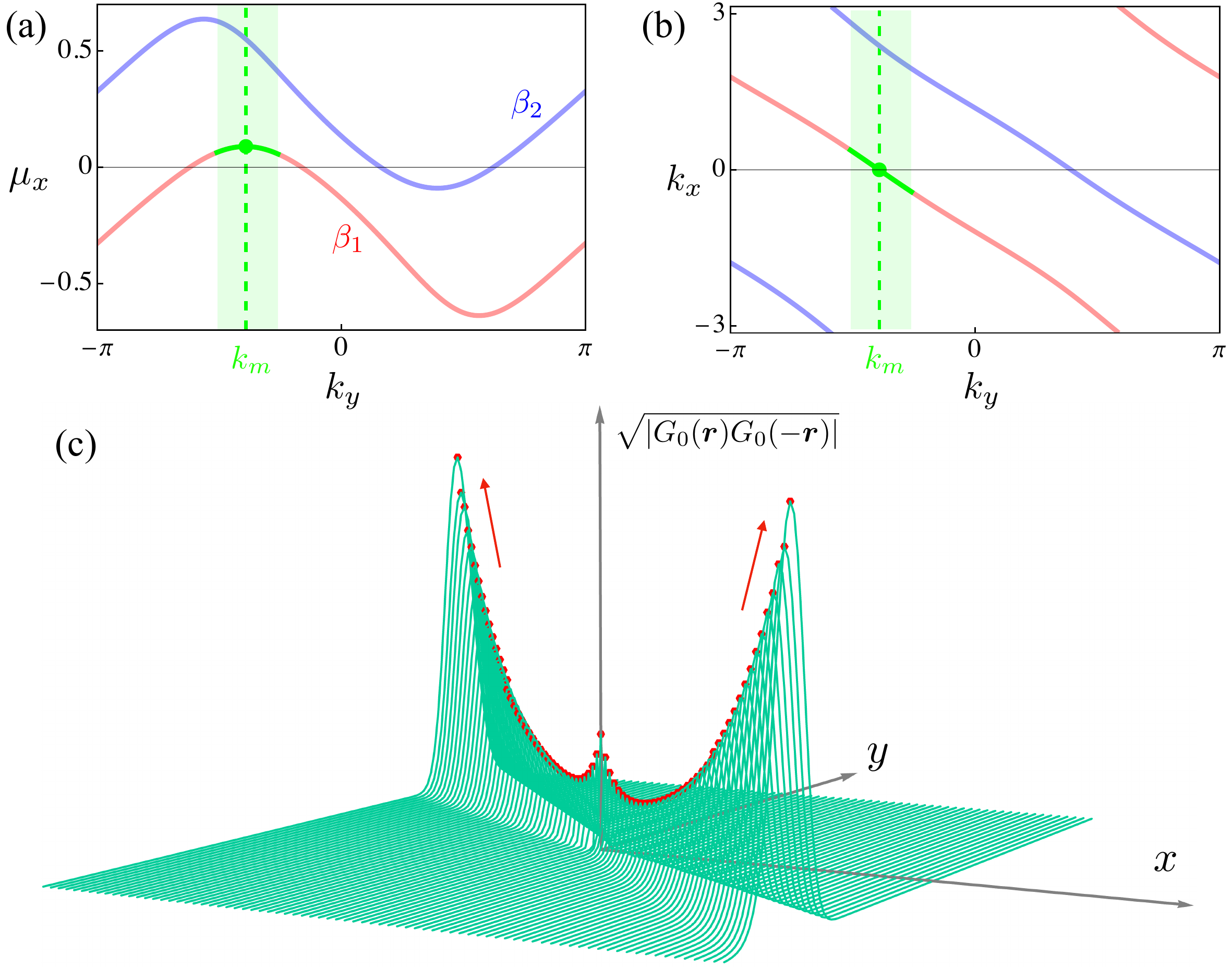}
    \caption{\textbf{The divergence of the forth-back propagator.} 
    (a)(b) are solution curves $\beta_1(k_y)$ and $\beta_2(k_y)$ of the characteristic equation $|\mathcal{H}_0(\beta_x,k_y)-E|=0$ where $\mu_x$ and $k_x$ are defined through $\beta_x=e^{\mu_x+i k_x}$, indicating the localization length and wavevector.
    $G_0(E;\mathbf{r}',\mathbf{r})$ is mainly contributed by the maximum of the $\beta_1$ at $k_y=k_m$, the vicinity of which is marked by green regions.
    (c) shows the divergence of the forth-back propagator $|G_0(\mathbf{r})G_0(-\mathbf{r})|$ in the $\mathbf{r}=(x,y)$ plane.
    }
    \label{fig:2}
\end{figure}

Now, we apply our formula to a reciprocal model that exhibits the algebraic skin effect, as given by Eq.~\eqref{eq:2DTBModel}. By introducing the non-Bloch wavevector $\beta_x := e^{i(k_x - i \mu_x)}$ and Bloch wavevector $k_y$, the Hamiltonian in Eq.~\eqref{eq:2DTBModel} can be reexpressed as $\mathcal{H}_0(\beta_x, k_y) = t_x (\beta_x + \beta_x^{-1}) + 2t_y \cos k_y + t_{xy} (\beta_x e^{ik_y} + \beta_x^{-1} e^{-ik_y}) + u$.
For an off-resonance $E$ and a $k_y \in [0, 2\pi)$, the characteristic equation $\mathcal{H}(\beta_x, k_y) = E$ yields two solutions for $\beta_x$, fully-gapped and ordered by their magnitudes as $|\beta_1(k_y)| < |\beta_2(k_y)|$ [see Fig.~\ref{fig:2}(a)].
Then, using the residue theorem, $G_0(E; \mathbf{r}_1,\mathbf{r}_2)$ can be calculated as (see Method I for details):
\begin{equation}\label{eq:real_space_prop2}
    G_0(E; \mathbf{r}_1,\mathbf{r}_2) = \int_{-\pi}^{\pi} \dd k_y \frac{[\beta_1(k_y)]^{x_1-x_2}}{\beta_1(k_y) - \beta_2(k_y)} e^{i k_y (y_1 - y_2)}.
\end{equation}
Due to reciprocity $H_0^T = H_0$, the correlation function is symmetric $G_0(E; \mathbf{r}_1,\mathbf{r}_2) = G_0(E; \mathbf{r}_2,\mathbf{r}_1)$, as shown in Fig.~\ref{fig:2}(c).

For large $L\equiv x_1-x_2$, if we properly select $\delta y \equiv y_1 - y_2$ such that the phase fluctuations in $[\beta_1(k_y)]^L$ and $e^{ik_y\delta y}$ are canceled, then this integral can be estimated by:
\begin{equation}
    |G_0(E; \mathbf{r}_1,\mathbf{r}_2)| = |G_0(E; \mathbf{r}_2,\mathbf{r}_1)| \sim e^{\mu_m L},
\end{equation}
where $\mu_m := \max_{k_y \in [0, 2\pi)} \ln{|\beta_1(k_y)|}>0$ [marked as the green dot in Fig.~\ref{fig:2} (a)]. 

It is worth noting that not all $\delta y$ will cause the integral in Eq.~\eqref{eq:real_space_prop2} to diverge as $L \rightarrow \infty$. We will address this key finding in the next section. In this case, the impurity threshold $V_*$ [Eq.~\eqref{eq:senscriterion}] to invalidate the perturbative expansion is obtained as $V_* \sim e^{-\mu_m L}$, where $\mu_m>0$. Consequently, the threshold $V_*$ tends to zero following an exponential law as $L \to \infty$ in the thermodynamic limit, expressed as:
\begin{equation}
    \lim_{L \rightarrow \infty} V_* \sim \lim_{L \rightarrow \infty} e^{-\mu_m L} \rightarrow 0.
\end{equation}
This indicates that even an infinitesimal bi-impurity strength can strongly alter the spectrum in the cylinder geometry. So far, we have explained the spectral sensitivity observed in Fig.~\ref{fig:1}.

Although the reciprocal example above is used to demonstrate the bi-impurity-induced ultra spectral sensitivity, this type of sensitivity remains robust and is commonly observed even when the system deviates from reciprocity. The non-reciprocity tends to suppress this type of sensitivity and the divergence of forth-back propagator. In the non-reciprocal limit, this sensitivity vanishes as the forth-back propagator converges. Remarkably, for off-resonance energies, the criterion in Eq.~\eqref{eq:senscriterion} holds true for any system.
See more discussions and examples in Supplementary Note III.

\section{The analytic phase diagram and the scale-free behavior under different thermodynamic limits}

\begin{figure}[t]
    \centering
    \includegraphics[width=\linewidth]{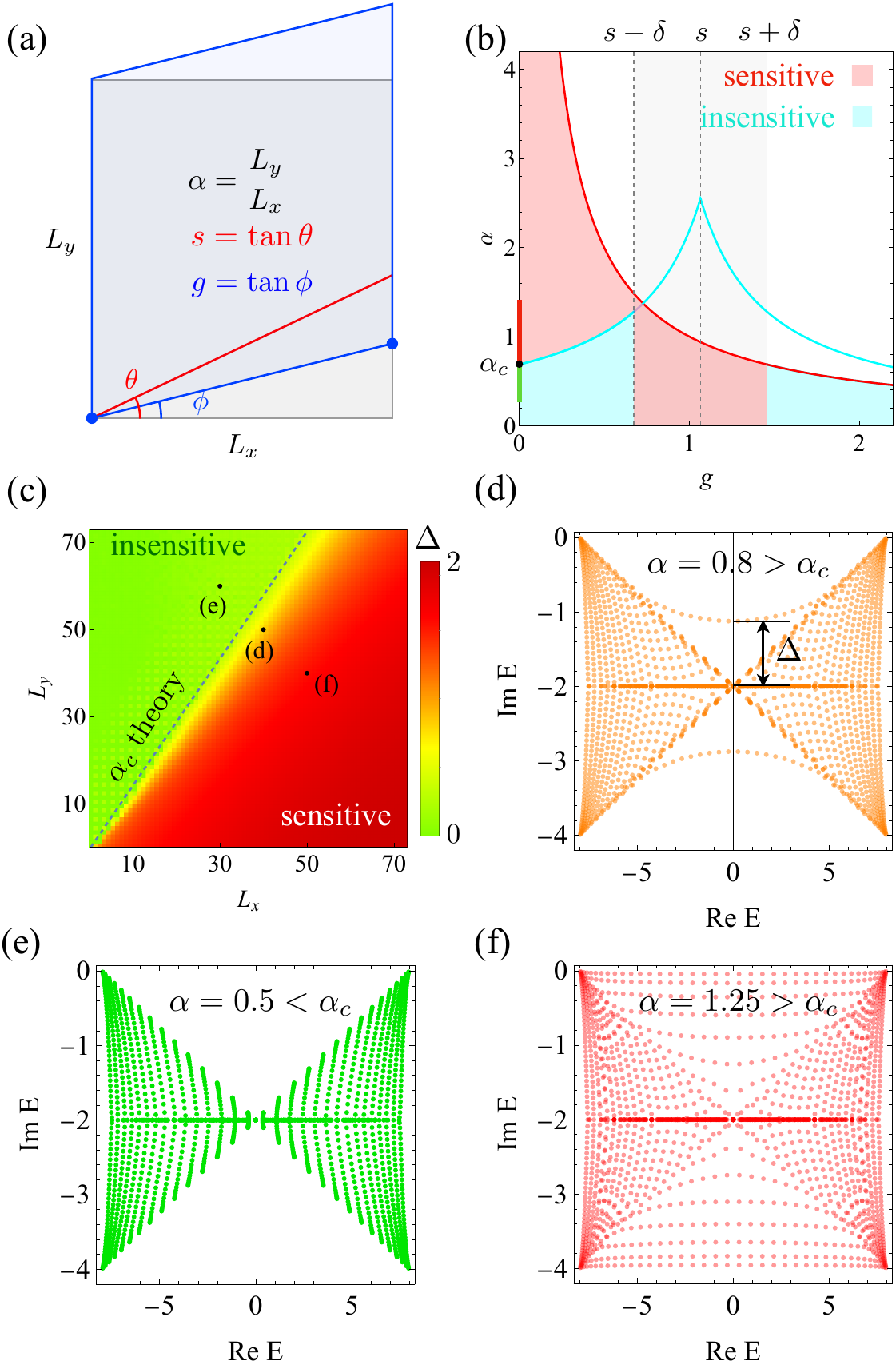}
    \caption{\textbf{Sensitive/insensitive phase diagram in the $(\alpha,g)$ space.}
    (a) Sketch shows the definitions of the geometry parameters. 
    The blue dots denote the locations of bi-impurities.
    The tilting angle is $\phi$ and $g=\tan\phi$.
    (b) The phase diagram shows the sensitive/insensitive region in the $\alpha-g$ plane.
    For fixed $g=0$, (c) shows the phase transition with different aspect ratios $\alpha$, which aligns with the phase diagram in (b). The order parameter $\Delta$ is defined in the OBC spectrum, as illustrated in (d). 
    (e) In the insensitive region ($\alpha<\alpha_c$), the spectral coverage is the same as the cylinder spectrum, indicating $\Delta=0$.
    (f) In the sensitive region ($\alpha>\alpha_c$), the OBC spectrum differs dramatically from the cylinder spectrum, which indicates the occurrence of spectral sensitivity and order parameter $\Delta\neq0$. }
    \label{fig:3}
\end{figure}

We have thus far demonstrated that the origin of spectral sensitivity lies in the diverging non-local response. 
However, the role of geometric factors: the spatial aspect ratio $\alpha$ and the bi-impurity location factor $g=\delta y/L$, remains unclear. 
In this section, we establish an analytic formula that generates the sensitivity phase diagram in the parameter space of $(\alpha,g)$ [Fig.~\ref{fig:3}(b)]. 
This phase diagram reveals that in the limit of large $L \rightarrow \infty$, the occurrence of spectral sensitivity is solely determined by the values of $(\alpha,g)$. 
Here, $\alpha$ represents different thermodynamic limits in two dimensions.
Since each pair $(\alpha,g)$ corresponds to a specific parallelogram geometry [as illustrated in Fig.~\ref{fig:3}(a)], the sensitivity phase diagram dictates the spectral coverage for each parallelogram under distinct thermodynamic limits [Figs.~\ref{fig:3}(c)-(f)]. 

In reciprocal systems, the divergence of the forth-back propagator coincides with that of $|G_0(E;\mathbf{r}_1,\mathbf{r}_2)|$. 
As shown in Eq.~\eqref{eq:real_space_prop2}, the integral is dominated by the maximum $\mu_m$ of the branch $\beta_1$, indicated by the green dot in Fig.~\ref{fig:2}(a). 
Near this maximum point, the branch $\beta_1(k_y)$ can be expanded as $\beta_1(k_y)\approx\exp{i [k_x(k_m)- s \delta k_y]+[\mu_m-c \, \delta k_y^2]}$, where the higher-order terms are neglected. 
Here, $s$ is the first-order derivative of $-k_x(k_y)$ with respect to $k_y$ at $k_y=k_m$; at the maximum $\mu_m$, the first derivative of the localization length function $\mu_x(k_y)$ vanishes, and its curvature is denoted by $-c$ ($c>0$). 
Substituting this expansion into Eq.~\eqref{eq:real_space_prop2}, neglecting constant phase factors, approximating the denominator as a constant near $k_y=k_m$, and discretizing the integral, we obtain (see  Method II for more detail):
\begin{align}
    G_0(E;L,\delta y) & \simeq e^{\mu_m L}\sum_{n=-\infty}^{\infty}e^{i 2\pi \alpha (g - s) n} e^{-\frac{4 \pi^2 \alpha^2 c}{L} n^2} \nonumber \\
    & = e^{\mu_m L} \vartheta_3\left(\pi \alpha (g-s);e^{- \frac{4 \pi^2 \alpha^2 c }{L}}\right), \label{eq:corelationfunc}
\end{align}
where $\alpha$ and $g$ are geometric parameters [Fig.~\ref{fig:3}(a)]; $s$ and $c$ describe the local behavior of the branch $\beta_1(k_y)$ near the maximum point $k_m$ [Figs.~\ref{fig:2}(a)(b)], which depend on the off-resonance energy $E$; and $\vartheta_3(z;q)$ is the Jacobi theta function~\cite{borwein1987pi}. 

For simplicity, we relabel $A=\pi \alpha (g-s)$ and $B=4\pi^2 \alpha^2 c > 0$. Since $\vartheta_3(z;q)$ is a periodic function of $z$ with a period of $\pi$, we can focus on the first periodicity where $0 \leq A < \pi$. 
At large $L$, the $\vartheta_3$ function can be expanded with respect to $1/L$ and Eq.~\eqref{eq:corelationfunc} becomes 
\begin{equation}\label{eq:phasediagram}
    \lim_{L\rightarrow \infty} G_0(E;L,\delta y) \sim
    \left\{  \begin{split}
    & e^{\mu_m L}e^{-\frac{A^2}{B}L}, \,\,\,\,\, 0 \leq A < \frac{\pi}{2}; \\  
    & e^{\mu_m L}e^{-\frac{(\pi-A)^2}{B}L}, \, \frac{\pi}{2} \leq A < \pi.
    \end{split}  \right.  
\end{equation}
The Eq.~\eqref{eq:phasediagram} provides complete information about the divergence of the forth-back propagator in the thermodynamic limit. 
For fixed geometry factors $\alpha$ and $g$, different energies $E$ lead to distinct values of $A,B$ and $\mu_m$. Consequently, the forth-back propagator diverges over certain regions of off-resonance energies. 

\begin{figure*}[t]
    \centering
    \includegraphics[width=\linewidth]{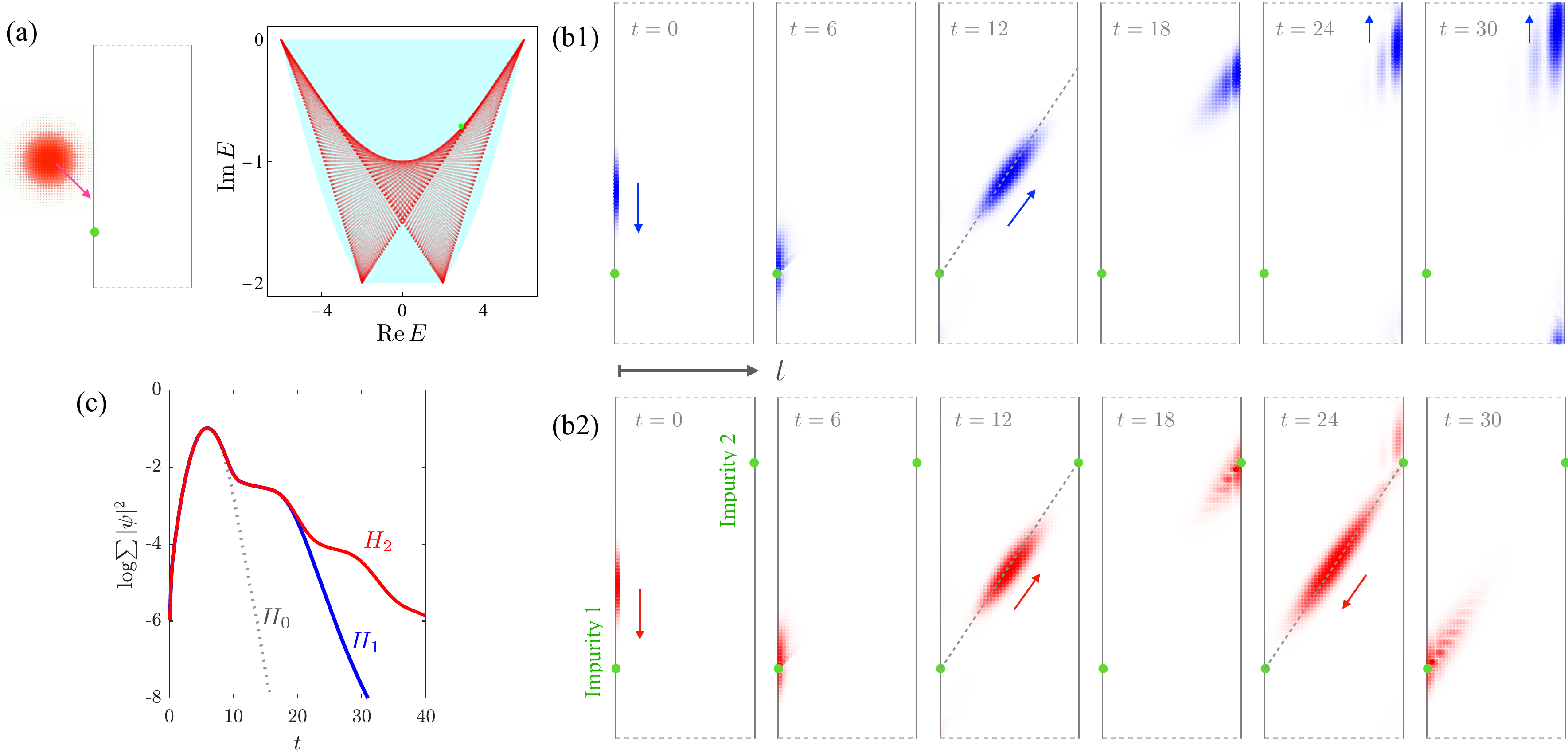}
    \caption{\textbf{Dynamical consequence of bi-impurity-induced spectral sensitivity.}
    (a) The sketch shows our setup: A wavepacket with frequency $\omega_0=2.89$ is injected from the left boundary and enters the non-Hermitian system, whose Hamiltonian is given by Eq.~\eqref{eq:2DTBModel} with parameters $t_x=2i,t_y=1,t_{xy}=2,u=-2i$.
    In our simulation, we take the cylinder geometry, and the system's spectrum is shown with a real frequency cut at $\Re E=\omega_0$ (indicated by the black line).
    Here, red dots represent cylinder eigenvalues, and the cyan region represents PBC spectral coverage.  
    The time slices of wavepacket dynamics evolution are compared between the single-impurity (b1) and double-impurity (b2) cases. 
    Before $t\simeq20$, the wavepackets behave similarly: they are scattered by the left impurity site into bulk waves.
    When the wavepacket reaches the right edge, if there is no second impurity (b1), the wavepacket will adhere to the edge and decay.
    If there is a second impurity potential (b2), the wavepacket will be scattered back as bulk wave until it is scattered again by the first impurity.
    (c) shows the logarithm of the total amplitude of the wavefunction included in the system. Comparison is provided between $H_0$, $H_1$, and $H_2$.
    Before $t\simeq5$, the total amplitude increases as the wavepacket is injected into the system.
    After that, the amplitude decays as the wave slides along the edge with a decay rate matching the green dot in (a). 
    When scattered by the first impurity, it becomes a bulk wave with a very low decay rate expanding into the cyan spectral region in (a).
    Consequently, a plateau can be observed in (c) for $H_1$ and $H_2$, while $H_0$ keeps decreasing.
    At $t\simeq20$, the wavepacket reaches the right edge. 
    For $H_2$, the second impurity again scatters the wavepacket back as a bulk wave following the same trajectory.
    This process repeats itself so that a series of plateaus in a stair-like shape can be observed (The red curve).
    In contrast, the amplitude of $H_1$ rapidly decays without a second plateau. }
    \label{fig:4}
\end{figure*}

Here, we consider the lower bound of energy $E$ at which the divergence first occurs. 
When $\mu_m-A^2/B>0$ or $\mu_m-(\pi-A)^2/B>0$, the spectral sensitivity can be induced by the bi-impurity potential, indicating that the system is in the sensitivity phase. 
Using this criterion, we can plot the sensitivity phase diagram, as illustrated in Fig.~\ref{fig:3}(b). 
Physically, the distance between two impurities $\delta y$ cannot exceed the length $L_y$, meaning $0 < \delta y < L_y = L/\alpha$. 
Therefore, we have $ 0 < g \alpha <1 $, as indicated by the red line, and the parameters are valid below this bound.  
Using Eq.~\eqref{eq:phasediagram}, one can finally obtain the phase diagram, with the red region for the sensitive phase and the cyan region for the insensitive phase shown in Fig.~\ref{fig:3}(b). 

From the formula Eq.~\eqref{eq:phasediagram}, we can draw the following quick conclusions:
(i)~When $g=s$, i.e., the two impurities are placed along the direction of $s$, it leads to $A=0$ and the forth-back propagator has to diverge, regardless of the aspect ratio $\alpha$. Consequently, the bi-impurity aligned along the direction of $s$ can always result in the spectral sensitivity as indicated in Fig.~\ref{fig:1}. 
(ii)~When $g$ is fixed, different $\alpha$ corresponds to distinct thermodynamic limits. As shown in Fig.~\ref{fig:3}(b), for example, when $g=0$, for some range of $\alpha < \alpha_c$, the system is in the insensitive phase, while for the range of $\alpha>\alpha_c$, the system is at the sensitive phase. It reveals the role of $\alpha$ in the spectral sensitivity. 

As a by-product of our formula [Eq.~\eqref{eq:phasediagram}], we find that the OBC spectrum converges to distinct coverage regions under different thermodynamic limits $\alpha$. 
Such a phenomenon was recently observed numerically in Ref.~\cite{HaiPing2024arXiv}. Here we provide an analytic explanation from the perspective of diverging non-local response. 
We start with a bi-impurity cylinder and link it to the OBC spectrum. 
If non-perturbative spectral sensitivity arises from adding two edge impurities, it will also occur for a line of identical impurities. This establishes a connection between the bi-impurity cylinder spectrum and the corresponding parallelogram geometry spectrum, where the geometry is defined by $\alpha$ and $g$, as shown in Fig.~\ref{fig:3}(a). 
In reciprocal systems, the notable difference in spectral coverage between fully OBC and cylinder geometries primarily results from this bi-impurity-induced sensitivity. 
Consequently, compared to the cylinder spectrum, the spectral coverage under OBC will differ significantly when the forth-back propagator diverges. 
Conversely, when the forth-back propagator converges, the OBC spectrum coverage is nearly unchanged and can be derived perturbatively from the cylinder spectrum. 
As a numerical verification, we fix $g = 0$ and calculate the OBC spectra for different geometric aspect ratios $\alpha$ [Figs.~\ref{fig:3}(d)(e)(f)]. 
We define the spectral shift $\Delta$ [Fig.~\ref{fig:3}(d)] as an order parameter to quantify the deviation from the cylinder-geometry spectrum.
As shown in Figs.~\ref{fig:3}(b)(c), when the $\alpha$ is below the critical value $\alpha_c$, the OBC spectrum exhibits the same coverage as that of the cylinder geometry [Fig.~\ref{fig:3}(e)]. 
In contrast, when $\alpha>\alpha_c$, the OBC spectrum undergoes a dramatic shift from the cylinder geometry spectrum [Fig.~\ref{fig:3}(f)]. 
Notably, the critical value of the aspect ratio $\alpha_c$ can be analytically determined by our formula in Eq.~\eqref{eq:phasediagram}, as indicated by the black dot marked in Fig.~\ref{fig:3}(b). 

\section{Dynamical consequence: wavepacket revival}

In this section, we explore the dynamical consequences of bi-impurity-induced sensitivity. 
The setup for the purpose is a heterojunction consisting of a Hermitian lead and the non-Hermitian bulk whose Hamiltonian is given by Eq.~\eqref{eq:2DTBModel} with parameters $t_x=i,t_y=1,t_{xy}=2,u=-2i$.
The size of the simulated system is $L_x\times L_y=82\times101$ and we take PBC in $y$ and OBC in $x$ directions.
Since the wavepacket dynamics is local and agnostic of distant boundary conditions, taking OBC/PBC in $y$ will have no major impact on the finite-time dynamical behaviors.
In Fig.~\ref{fig:4} (a), the sketch shows the basic setup of the dynamical simulation as well as the energy spectrum of the system.
At time $t=0$, a Gaussian wavepacket with real frequency $\omega_0=2.89$ and half-width $\sigma=7$ is injected into the non-Hermitian system from the left lead.
The wavepacket time evolution with a single impurity (the upper panels) and bi-impurity cases (the lower panels) are compared in Fig.~\ref{fig:1}(b). 
When the wavepacket enters the system and touches the left edge, it effectively feels the cylinder geometry spectrum [the red dots in Fig.~\ref{fig:4}(a)]. 
After a while, the dominant component is the one with the largest imaginary part [marked as the green dot in the spectrum], which behaves as edge states of the cylinder-geometry system. 
Therefore, in the first two slices of the dynamics, the wavepacket evolves along the boundary until it hits the first impurity site (the green dot).
The wavepacket is scattered into the bulk and released from the cylinder spectrum. The scatted direction can be determined by the equal frequency contour in momentum space. 
The decay rate of the wavepacket varies with time, as shown in Fig.~\ref{fig:4} (c), where we compute the total wavepacket amplitude over the whole system.
Before hitting the first impurity, the decay rate of both $H_1$ and $H_2$ agrees with the maximum imaginary of cylinder spectrum at $\omega_0$, as shown in Fig.\ref{fig:4} (c).
However, after being scattered by the first impurity, the slope suddenly levels out which aligns with our expectation.
Until now, the dynamical behaviors of $H_1$ and $H_2$ have been the same.

The difference shows up when the wavepacket reaches the right edge.
For $H_1$, there is no impurity side on the right boundary.
Therefore, the wavepacket again decays with a larger decay rate while gliding along the right edge.
For $H_2$, nevertheless, the wavepacket is scattered by the second impurity dot and again bounced back into the bulk with a low decay rate.
After that, the wavepacket will again hit the left edge and bounce back and forth. 
Therefore, the two edge impurities establish a long-lived scattering channel that allows the wavepacket bounces forth and back between them with a low decay, which we term ``wavepacket revival". 
In Fig.~\ref{fig:4}(c), we can observe a series of plateaus in a staircase shape, as opposed to only one plateau for $H_1$.
It is illuminating to compare this simulation to Hermitian systems or non-reciprocal non-Hermitian systems with line skin effects. 
For line skin effect, the wavepacket will be confined at the left edge, regardless of the edge impurities. 
While for the system exhibiting the bi-impurity spectral sensitivity, the two edge impurities build a long-distance communication channel such that the wavepacket travels forth and back between two edges. 
The core physics behind this lies in the bi-impurity-induced wavefunction, which exhibits quasi-long-range algebraic decay, enabling a non-local response between two distant impurities. 

\section{Methods}

\appendix

\section{Method I: Divergence of forth-back propagator}\label{MethodI}

First, we show the derivation of the spectral radius of the bi-impurity response operator, given by Eq.\eqref{eq:biimp_response}.
In the real-space basis, the bi-impurity potential $\hat{V}$ and $\hat{G}_0$ is represented by
\begin{align}
    \hat{V}&=V_1\delta_{\mathbf{r},\mathbf{r}_1}+V_2\delta_{\mathbf{r},\mathbf{r}_2},\nonumber\\
    \hat{G}_0&=\sum_{i,j=1}^N G_0(E;\mathbf{r}_i,\mathbf{r}_j)\delta_{\mathbf{r}_i,\mathbf{r}_j}.
\end{align}
Therefore, the response operator is given by
\begin{equation}
    \hat{G}_{0}\hat{V}{=}V_1\sum_{i=1}^{N}G_0(E;\mathbf{r}_i,\mathbf{r}_1)\delta_{\mathbf{r}_i,\mathbf{r}_1}\nonumber{+}V_2\sum_{i=1}^{N}G_0(E;\mathbf{r}_i,\mathbf{r}_2)\delta_{\mathbf{r}_i,\mathbf{r}_2}.
\end{equation}
Properly rearranging the real-space basis such that $\mathbf{r}_1$ and $\mathbf{r}_2$ are listed as the first two, then written as a matrix, the response operator can be represented as:
\begin{equation}
\hat{G}_0\hat{V}=
    \begin{pmatrix}
        V_1 G_{0}(E;\mathbf{r}_1,\mathbf{r}_1) & V_2 G_{0}(E;\mathbf{r}_1,\mathbf{r}_2) & 0 & \cdots  & 0\\
        V_1 G_{0}(E;\mathbf{r}_2,\mathbf{r}_1) & V_2 G_{0}(E;\mathbf{r}_2,\mathbf{r}_2) & 0 & \cdots  & 0\\
        \vdots  & \vdots  & \vdots  & \ddots  & \vdots \\
        V_1 G_{0}(E;\mathbf{r}_N,\mathbf{r}_1) & V_2 G_{0}(E;\mathbf{r}_N,\mathbf{r}_2) & 0 & \cdots  & 0
    \end{pmatrix}.
\end{equation}
The eigenvalues of this matrix consist of $(N-2)$ zeros and the two non-zero eigenvalues of the top-left submatrix. Therefore, the spectral radius $\rho(\hat{G}_0\hat{V})$ is determined by the spectral radius of the top-left $2\times 2$ submatrix.
\begin{equation}
     \rho (\hat{G}_0\hat{V}) = \rho
    \mqty(V_1 G_{0}(E;\mathbf{r}_1,\mathbf{r}_1) & V_2 G_{0}(E;\mathbf{r}_1,\mathbf{r}_2) \\ V_1 G_{0}(E;\mathbf{r}_2,\mathbf{r}_1) & V_2 G_{0}(E;\mathbf{r}_2,\mathbf{r}_2)).
\end{equation}
It is worth noting that although we have utilized onsite impurity here as $\hat{V}$, using impurity bonds also works (e.g. magnetic flux through a plaquette).
In fact, as long as the impurity is a local disorder and the size of the disorder tends to zero compared to the system size in the thermodynamic limit, the preceding derivation still applies, where $\mathbf{r}_1$ and $\mathbf{r}_2$ represent the locations of the bi-impurities.

Next, we derive the integral representation of the Green's function in Eq. (4) using the residue theorem. According to the definition, we have
\begin{align}
G_{0}( E;\mathbf{r}_1,\mathbf{r}_2) & =\mel{\mathbf{r}_1}{( E\hat{I} -\hat{H}_{0})^{-1}}{\mathbf{r}_2}\\
 & =\int _{0}^{2\pi }\mathrm{d} k_{y}\oint _{\beta _{x} \in \text{GBZ}}\frac{\mathrm{d} \beta _{x}}{\beta _{x}}\frac{e^{ik_{y}( y_1-y_2)} \beta _{x}^{x_1-x_2}}{E-\mathcal{H}_{0}( \beta _{x} ,k_{y})}.\nonumber
\end{align}
where we have dropped the constant prefactor as will be done in the following derivations. For simplicity, we consider nearest-neighbor-hopping models, and the generalized deduction will be shown in the Supplementary Note II.1.
The characteristic equation in the $k$-space can be written as:
\begin{equation}
E-\mathcal{H}_{0}( \beta _{x} ,k_{y}) =\frac{( \beta _{x} -\beta _{1}( k_{y}))( \beta _{x} -\beta _{2}( k_{y}))}{\beta _{x}}.
\end{equation}
Here, as in the main text, we have defined the two roots of the characteristic equation as $\beta _{1}$ and $\beta _{2}$ satisfying $|\beta _{2}( k_{y}) | >|\beta _{1}( k_{y}) |$.
When $E\in\sigma_{\text{cyl}}$, the standing wave condition is given by $|\beta_1|=|\beta_2|$. Now we consider off-resonance $E$, then the two solutions are fully gapped and the cylinder GBZ curve is between these two solutions. We can use these two solutions to rewrite the Green's function as the following
\begin{align}
G_{0}( E;\mathbf{r}_1,&\mathbf{r}_2)=\int _{0}^{2\pi }\mathrm{d} k_{y}\, e^{ik_{y}( y_1-y_2)} \times \\
&\oint _{\beta _{x} \in \text{GBZ}}\mathrm{d} \beta _{x}\frac{\beta _{x}^{x_1-x_2}}{( \beta _{x} -\beta _{1}( k_{y}))( \beta _{x} -\beta _{2}( k_{y}))}.\nonumber
\end{align}
\begin{figure}[h]
    \centering
    \includegraphics[width=0.9\linewidth]{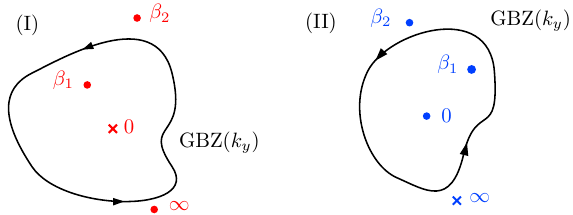}
    \caption{The distribution of isolated singularities in the two cases. The dots denote poles while the crosses represent zeros of the integrand.}
    \label{fig:poles}
\end{figure}
We first perform the contour integral over $\beta_x$ along the GBZ curve. Without loss of generality, assume $x_1>x_2$, then the integrand has three poles: $\beta_1$, $\beta_2$, and $\infty$, where $\beta_1$ is inside the GBZ curve while the rest are outside [see Fig.~\ref{fig:poles} (I)]. 
Using the residue theorem, we get
\begin{equation}
G_{0}( E;\mathbf{r}_1,\mathbf{r}_2){=}\int _{0}^{2\pi }\mathrm{d} k_{y}\frac{[ \beta _{1}( k_{y})]^{x_1-x_2}e^{ik_{y}(y_1-y_2)}}{\beta _{1}( k_{y}) -\beta _{2}( k_{y})}.
\end{equation}
Similarly, when we consider $G_0(E;\mathbf{r}_2,\mathbf{r}_1)$, we have
\begin{align}
G_{0}(&E;\mathbf{r}_2,\mathbf{r}_1)=\int _{0}^{2\pi }\mathrm{d}\, k_{y} e^{ik_{y}( y_2-y_1)} \times \\
&\oint _{\beta _{x} \in \text{GBZ}}\mathrm{d} \beta _{x}\frac{1}{\beta _{x}^{x_1-x_2}( \beta _{x} -\beta _{1}( k_{y}))( \beta _{x} -\beta _{2}( k_{y}))}.\nonumber
\end{align}
Now the integrated has three poles: $\beta_1$, $\beta_2$, and $0$ [see Fig.~\ref{fig:poles} (II)]. 
Note that the $\beta_x=0$ pole is of order $|x_1-x_2|$, which is difficult to evaluate.
Since the residues of all poles on the complex plane add up to $0$, we can instead only consider the pole $\beta_2$ outside the loop (note that $\beta_x=\infty$ is no longer a pole now).
Hence,
\begin{equation}
G_{0}( E;\mathbf{r}_2,\mathbf{r}_1) =\int _{0}^{2\pi }\mathrm{d} k_{y}\,\frac{[ \beta _{2}( k_{y})]^{x_2-x_1}e^{ik_{y}( y_2-y_1)}}{\beta _{1}( k_{y}) -\beta _{2}( k_{y})}.
\end{equation}
Next, we perform approximation for $G_0(E;\mathbf{r}_1,\mathbf{r}_2)$ as we approach the thermodynamic limit. 
Note that the value of this integral is mainly contributed by the integrand in the vicinity of $k_{y}=k_{m}$.
We can first omit the denominator as its variation is slow without being exponentiated by the system size. 
Thus,
\begin{equation}
G_{0}( E;\mathbf{r}_1,\mathbf{r}_2) \simeq \int _{0}^{2\pi }\mathrm{d} k_{y}\,[ \beta _{1}( k_{y})]^{x_1-x_2} e^{ik_{y}( y_1-y_2)}.
\end{equation}
Since this integral is mainly contributed by the maximum value of $\beta _{x}( k_{m})$ and its vicinity, we first expand the $\beta _{x}( k_{y})$ around $k_{y} =k_{m}$ as
\begin{equation}
\beta _{x}( k_{y}) =\exp[\mu _{m} -is( k_{y} -k_{m}) -c( k_{y} -k_{m})^{2}] .
\end{equation}
Here we have both expanded the real and imaginary part of $\log \beta _{x}( k_{y})$ to the leading order. The first-order expansion of the real part is zero. Here, $\mu _{m} =\log |\beta _{x}( k_{m}) |$, $s$ is the rate of variation of the phase at $k_{m}$, and $-c$ is the curvature of the $\mu ( k_{y})$ curve at $k_{y} =k_{m}$. Next, to let $L_{y}$ enter the picture, we discretize the integral $\delta k=k_{y} -k_{m} =\frac{2\pi n}{L_{y}}$ with $n\in \mathbb{Z}$, and then take the expansion into the expression and get:
\begin{align}
G_{0}( E;\mathbf{r}_1,\mathbf{r}_2) & \simeq e^{\mu _{m} \delta x}\times\\ &\sum _{n\in \mathbb{Z}}\exp\left( i\frac{2\pi n}{L_{y}}( \delta y-s\delta x) -c\left(\frac{2\pi n}{L_{y}}\right)^{2} \delta x\right).\nonumber
\end{align}
According to the definition of the Jacobi Theta function, $\vartheta _{3}( z,q) =\sum _{n\in \mathbb{Z}} e^{2niz} q^{n^{2}}$, the current approximated expression can be reformulated as
\begin{equation}
G_{0}( E;\mathbf{r}_1,\mathbf{r}_2){\simeq}e^{\mu _{m} \delta x} \vartheta _{3}\left(\frac{\pi }{L_{y}}( \delta y-s\delta x) ,e^{-\frac{4\pi ^{2} c\delta x}{L_{y}^{2}}}\right).
\end{equation}
For the bi-impurity cylinder, we take $\delta x=L_x$. Set $\alpha=L_x/L_y$, $g=\delta y/L_x$, and $L_x=L$, we obtain
\begin{equation}\label{eq:theta}
    G_0(E;\mathbf{r}_1,\mathbf{r}_2)\simeq e^{\mu_m L}\vartheta_3\left(\pi\alpha|g-s|,e^{-4\pi^2 c\alpha^2\frac{1}{L}}\right).
\end{equation}
Note that $\vartheta_3(z,q)=\vartheta_3(-z,q)$, therefore, $g-s$ can be written as $|g-s|$. 

\section{Method II: Sensitivity Phase Diagram}
In this method section, given the geometric parameters $(g,\alpha)$, we study when Eq.~\eqref{eq:theta} diverges in the thermodynamic limit $L\rightarrow \infty$. 
Subsequently, we can obtain the phase diagram Fig.~\ref{fig:3} (a).
Notice that $\alpha < 1/g$ due to the constraint $\delta y< L_{y}$ and $\alpha $ has a period of $1/|g-s|$ in $\vartheta _{3}$. We focus on the first period. Now define $\epsilon =1/L$, we want to study the limit
\begin{equation}
\begin{aligned}
\lim _{L\rightarrow +\infty} G_{0}( L) & =\lim _{\epsilon \rightarrow 0^{+}} e^{\frac{\mu _{m}}{\epsilon }} \vartheta _{3}\left( A( \alpha ) ,e^{-B( \alpha ) \epsilon }\right)\\
 & \sim \lim _{L\rightarrow +\infty } e^{C( \alpha ) L}
\end{aligned}
\end{equation}
where $A( \alpha ) =\pi \alpha |g-s|$, $B( \alpha ) =4\pi ^{2} c\alpha ^{2}$ and the sign of $C( \alpha )$ tells us if this expression is divergent. \ This will give us a range of $\alpha $ where $G_{0}$ is divergent. As this range is energy-dependent, we must exhaust all the off-resonance energies. 

The asymptotic expansion for the $\vartheta _{3}$ function as $\epsilon \rightarrow 0$ gives
\begin{equation}
\vartheta _{3}\left( A,e^{-B\epsilon }\right) \sim R( \epsilon ) e^{-\frac{A^{2}}{B\epsilon }}\left( e^{\frac{\pi ( 2A-\pi )}{B\epsilon }} +1\right)
\end{equation}
where $R( \epsilon )  >0$ is a rational function of $\epsilon $ resulting in a power law modification that won't affect the exponential behavior in the thermodynamic limit.
Note this expansion works when $0<A<\pi$.
Now we have two cases depending on the sign of $( 2A-\pi )$:

\noindent (I) $2A< \pi $ or equivalently $0< \alpha < 1/( 2|g-s|)$. Then, we have
\begin{equation}
\vartheta _{3}\left( A,e^{-B\epsilon }\right) \sim R( \epsilon ) e^{-\frac{A^{2}}{B\epsilon }} .
\end{equation}
Therefore,
\begin{equation}
C( \alpha ) =\frac{\mu _{m} B( \alpha ) -A( \alpha )^{2}}{B( \alpha )} =\frac{\pi ^{2} \alpha ^{2}\left( 4\mu _{m} c-( g-s)^{2}\right)}{B( \alpha )} .
\end{equation}
Since $\pi ^{2} \alpha ^{2} /B>0$, there are again two cases depending of the sign of $\left( 4\mu _{m} c-( g-s)^{2}\right)$:

(I.1) If $2\sqrt{\mu _{m} c} < |g-s|$, $C( \alpha ) < 0$ and $G_{0}$ converges.

(I.2) If $2\sqrt{\mu _{m} c} \geqslant |g-s|$, $C( \alpha ) \geqslant 0$ and $G_{0}$ diverges.

\noindent (II) $\pi\leq 2A<2\pi$ or equivalently $\alpha \geqslant 1/( 2|g-s|)$. Then, we have
\begin{equation}
\vartheta _{3}\left( A,e^{-B\epsilon }\right) \sim R( \epsilon ) e^{-\frac{( \pi -A)^{2}}{B\epsilon }} .
\end{equation}
Therefore,
\begin{equation}
\begin{aligned}
C( \alpha ) & =\frac{\mu _{m} B( \alpha ) -( \pi -A( \alpha ))^{2}}{B( \alpha )}\\
 & =\frac{\pi ^{2}}{B( \alpha )}\left( 4\mu _{m} c-|g-s|^{2}\right)( \alpha -\alpha _{+})( \alpha -\alpha _{-})
\end{aligned}
\end{equation}
where
\begin{equation}
\alpha _{\pm } =\frac{1}{|g-s|\pm 2\sqrt{\mu _{m} c}}
\end{equation}
Now $C( \alpha )$ is a quadratic equation of $\alpha $ with two roots $\alpha _{\pm }$. Based on the sign of $\left( 4\mu _{m} c-|g-s|^{2}\right)$ we again have two cases:

(II.1) If $2\sqrt{\mu _{m} c} \geqslant |g-s|$, then the roots $\alpha _{\pm }$ have the following order $\alpha _{+}  >0 >\alpha _{-}$. Moreover, $C( \alpha )$ is negative when $0< \alpha < \alpha _{+}$ and it is non-negative when $\alpha \geqslant \alpha _{+}$. However, $\alpha \geqslant 1/( 2|g-s|)  >\alpha _{+}$. Thus, $C( \alpha )  >0$ and $G_{0}$ diverges. This can be combined with the case (I.2) and we get the following conclusion: As long as $|g-s|\leqslant 2\sqrt{\mu _{m} c}$, $G_{0}$ always diverges. The physical meaning is that, as long as the bi-impurity direction is close enough to the maximal amplifying direction $s=-\partial_{k_y}k_x(k_m)$, the Green's function must be divergent.

(II.2) If $2\sqrt{\mu _{m} c} < |g-s|$, then the roots $\alpha _{\pm }$ have the following order $\alpha _{-}  >\alpha _{+}  >0$. Then $C( \alpha )$ is positive only when $\alpha _{+} < \alpha < \alpha _{-}$. Since in this case, we have $1/( 2|g-s|) \leqslant \alpha < 1/|g-s|$, then we can prove that $\alpha _{+} < 1/( 2|g-s|)$ and $\alpha _{-}  >1/|g-s|$. 
Consequently, $C( \alpha )  >0$ only when $\alpha  >\alpha _{+}$, i.e., the critical value of the aspect ratio $\alpha_c$ in the phase diagram is given by
\begin{equation}
\alpha  >\alpha _{c} =\alpha _{+} =\frac{1}{|g-s|+2\sqrt{\mu _{m} c}}.
\end{equation}

\bibliography{refs.bib}

\appendix
\onecolumngrid
\newpage
\begin{center}
    \textbf{Supplementary Information}
\end{center}
\section{I.~Numerical observation of geometry dependence}
\begin{figure}[h]
    \centering
    \includegraphics[width=\linewidth]{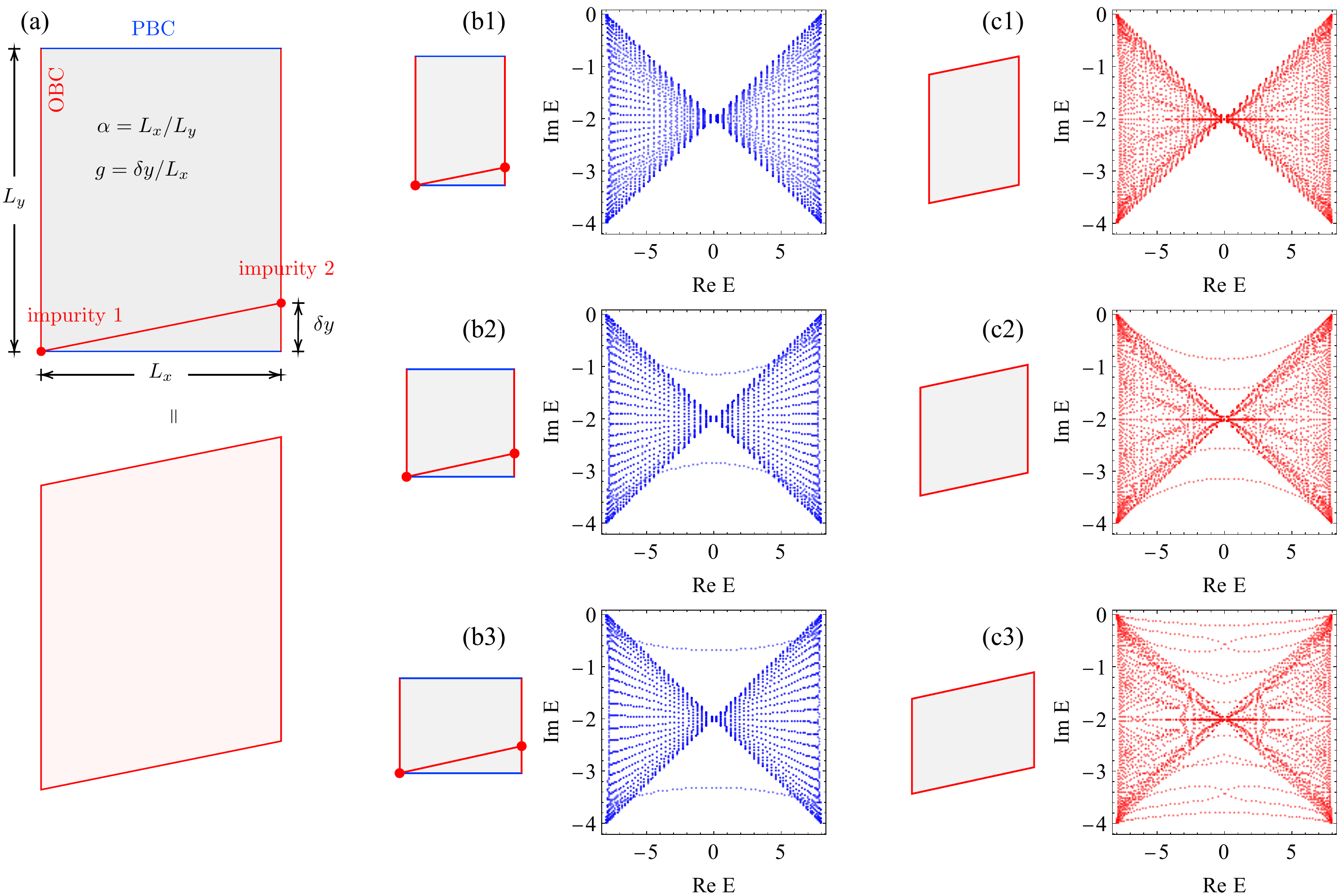}
    \caption{\textbf{The spectral sensitivity under different aspect ratio $\alpha$, for the same model with a fixed $g$.}
    (a) defines the geometric parameters $g$ and $\alpha$, and graphically shows that $(\alpha,g)$ uniquely defines a parallelogram shape up to scaling.
    (b1)-(b3) shows the spectrum of bi-impurity cylinder with the same $g$ but different $\alpha$: $\alpha=0.7,1.0,$ and $1.3$, respectively.
    (c1)-(c3) shows the spectrum of the corresponding parallelogram.
    }
    \label{fig:sf1}
\end{figure}

In the main text, we claim that the sensitivity depends on two critical geometric factors: $g=\delta y/L_x$ and $\alpha=L_x/L_y$.
These two parameters' definitions are shown in Fig.~\ref{fig:sf1}(a).
In fact, these two parameters uniquely determine a parallelogram (up to scaling).
Here, we provide some numerical observations showcasing the dependence of spectral sensitivity on these geometric factors.

In Fig.~\ref{fig:3} of the main text, we fix $g=0$ and change the aspect ratio $\alpha$.
We consider a more generic example with a fixed $g=0.2$ and altering $\alpha_1=0.7$, $\alpha_2=1.0$, and $\alpha_3=1.3$.
The model used is still given by Eq.~\eqref{eq:2DTBModel} with parameters $(t_x,t_y,t_{xy},u)=(i,0,4,-2i)$.
We compare two settings: one is the bi-impurity cylinder setting marked in blue, while the other is the corresponding parallelogram OBC marked in red.
According to the analytical approach in the main text, the threshold $\alpha_c$ for the aspect ratio for our model is given by
\begin{equation}
    \alpha_c=\frac{1}{|g-s|+2\sqrt{c\mu_m}}\simeq 0.795
\end{equation}
In Fig.~\ref{fig:sf1}(b1)-(b3) corresponding to $\alpha_1<\alpha_c$, $\alpha_2>\alpha_c$, and $\alpha_3>\alpha_c$ respectively, (b2) and (b3) shows spectral lines beyond the cylinder spectrum induced by the spectral sensitivity while (b1) does not, fully agree with our prediction.
Similarly, for Fig.~\ref{fig:sf1}(c1)-(c3), (c2) and (c3) shows the spectral sensitivity while (b1) does not.

In the next example, we fix the aspect ratio $\alpha=1$ and only change the relative placement of the impurities $g_1=0.3$, $g_2=0.6$, and $g_3=0.9$.
According to the phase diagram, $g_1$ is beyond the critical curve $\alpha(g)=1/(g-s+2\sqrt{c\mu_m})$.
\begin{figure}[h]
    \centering
    \includegraphics[width=\linewidth]{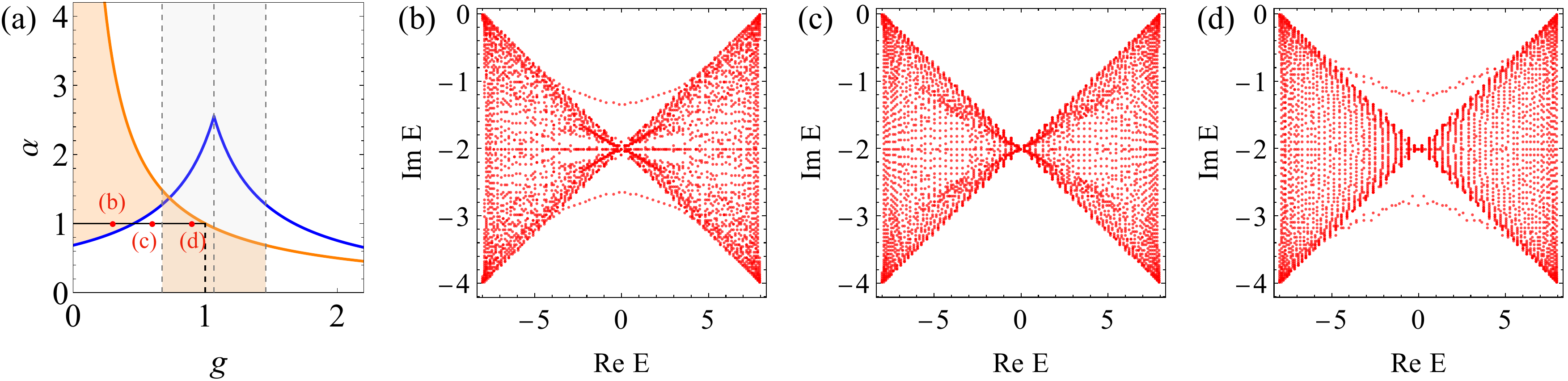}
    \caption{\textbf{The spectral sensitivity under different $g$, for the same aspect ratio $\alpha$.}
    (a) is the analytically obtained phase diagram, where the orange region is the sensitive region.
    The fixed $\alpha=1$ is marked as the solid black line, on which there are three red dots corresponding to the system (b), (c), and (d), with $g_1=0.3$, $g_2=0.6$, and $g_3=0.9$.
    The spectra are shown in red dots in (b), (c), and (d), where (c) is insensitive and (b) and (d) are sensitive.
    }
    \label{fig:sf2}
\end{figure}
Both $g_2$ and $g_3$ are below this threshold. 
However, $g_3$ is very close to the skewness direction, or more rigorously, $|g-s|<2\sqrt{c\mu_m}$.
Therefore, only $g_2$ is insensitive and both $g_1$ and $g_3$ are sensitive.
The numerical calculation is shown in Fig.~\ref{fig:sf2} which agrees with the theoretical phase diagram.

\section{II.~Generalizations}
Here, we discuss the validity of our theory in the general setting.
In subsection II.1, we derive the general expression for the forth-back propagator for long-range, non-reciprocal models.
In subsection II.2, we define the maximally sensitive region for the reciprocal models and prove that it is equivalent to the coverage of the PBC spectrum.

\subsection{II.1~Derivation of the forth-back propagator for general models}
In this subsection, we show that the spectral sensitivity is controlled by the two $\beta$ solutions closest to the cylinder GBZ for a generic model with long-range hopping.
When deriving Eq.\eqref{eq:real_space_prop2} in the Method I section, we assume the horizontal hopping range of the model is 1, so that the characteristic equation $\det|E\mathbb{I}-\mathcal{H}_0(\beta_x,k_y)|=0$ only has two solutions $\beta_1$ and $\beta_2$.

Here, we consider the following general model with the horizontal hopping range larger than 1 so that there are multiple solutions:
\begin{equation}
    \begin{aligned}
        \mathcal{H}_{0}( \beta _{x} ,k_{y}) & =t_{N}( k_{y}) \beta _{x}^{N} +\cdots +t_{-M}( k_{y}) \beta _{x}^{-M}.
    \end{aligned}
\end{equation}
For an given complex energy $E\notin \sigma _{\text{cyl}}$ outside the spectral region of $\mathcal{H}_0$ the characteristic equation reads
\begin{equation}
\det |E\mathbb{I} -\mathcal{H}_{0}( \beta _{x} ,k_{y}) |=\frac{( \beta _{x} -\beta _{M+N}( k_{y})) \cdots ( \beta _{x} -\beta _{1}( k_{y}))}{\beta _{x}^{M}}
\end{equation}
where $|\beta _{1}( k_{y}) |\leqslant \cdots \leqslant |\beta _{M+N}( k_{y}) |$ are the $M+N$ solutions ordered by their magnitudes. According to the 1D GBZ theory, the standing wave condition is given by $|\beta _{M}( k_{y}) |=|\beta _{M+1}( k_{y}) |$. Now $E$ is not an eigenvalue of $\mathcal{H}_{0}$, so the GBZ curve is between $\beta _{M}$ and $\beta _{M+1}$.
In the following, we show that the divergence of the forth-back propagator is controlled by the extrema of the two solution branches closest to the GBZ, i.e., the $\beta_M$ and $\beta_{M+1}$ branches.
When evaluating the matrix element $G_{0}( E;\mathbf{r},\mathbf{r} ')$, we have
\begin{equation}
G_{0}( E;\mathbf{r},\mathbf{r} ') =\int \mathrm{d} k_{y}\oint _{\beta _{x} \in \text{GBZ}}\frac{\mathrm{d} \beta _{x}}{\beta _{x}}\frac{\beta _{x}^{M+x-x'} e^{ik_{y}( y-y')}}{( \beta _{x} -\beta _{M+N}( k_{y})) \cdots ( \beta _{x} -\beta _{1}( k_{y}))}.
\end{equation}
Without loss of generality, we first consider when $x >x'$ and denote $\mathbf{r},\mathbf{r} '=( \delta x,\delta y)$. Then, for a fixed $k_{y}$, the poles that are included within the contour are $\beta _{1} ,\cdots ,\beta _{M}$. Using residue theorem, we get
\begin{equation*}
G_{0}( E;\mathbf{r},\mathbf{r} ') =\sum _{j=1}^{M}\int _{-\pi }^{\pi }\mathrm{d} k_{y}\frac{[ \beta _{j}( k_{y})]^{M-1+\delta x} e^{ik_{y} \delta y}}{( \beta _{j}( k_{y}) -\beta _{M+N}( k_{y})) \cdots ( \beta _{j}( k_{y}) -\beta _{1}( k_{y}))}.
\end{equation*}
Here, we still denote $\beta _{j}( k_{y}) =\exp( \mu _{j}( k_{y}) +ik_{j}( k_{y}))$. When $\delta x$ is large, the dominant term is the one proportional to $\beta _{M}^{M-1+\delta x}$ and the dominant contribution comes from the vicinity of the maximum of the $\mu _{M}( k_{y})$ curve at $k_{y} =k_{m}$. 
Notice that the denominator does not scale with system size.
Therefore, in the thermodynamic limit, it can be omitted and we get
\begin{equation}\label{eq:sup_G0approx1}
    G_0(E;\mathbf{r},\mathbf{r}')\simeq\int_{-\pi}^{\pi}\dd k_y[\beta_M(k_y)]^{M-1+\delta x}e^{i k_y\delta y}.
\end{equation}
We first expand the $\beta_M(k_y)$ near $k_y=k_m$ as
\begin{equation}
    \beta_M(k_y)=\exp[\mu_{\text{max},M}-is_M(k_y-k_m)-c_M(k_y-k_m)^2].
\end{equation}
Take this expansion into Eq.~\eqref{eq:sup_G0approx1}, we get
\begin{equation}
\begin{aligned}
    |G_0(E;\mathbf{r},\mathbf{r}')|&\simeq e^{(M-1+\delta x)\mu_{\text{max},M}}\int_{-\infty}^{\infty}\dd k_y\, e^{-is_M k_y(M-1+\delta x)-c_M k_y^2(M-1+\delta x)+ik_y\delta y}\\
    &\simeq e^{(M-1+\delta x)\mu_{\text{max},M}}\sum_{n=-\infty}^{\infty}\exp[i (s_M\delta x-\delta y+s_M (M-1))\cdot\left(\frac{2\pi n}{L_y}\right)-(c_M\delta x+c_M (M-1))\left(\frac{2\pi n}{L_y}\right)^2].
\end{aligned}
\end{equation}
In the thermodynamic limit, $|M-1|\ll L_y$, $|M-1|\ll L_y$, so we also neglect all the terms involving $(M-1)$ and using the definition of the geometric factors $\alpha=L_x/L_y=\delta x/L_y$, $g=\delta y/L_x$ and finally we arrive at
\begin{equation}
\begin{aligned}
    |G_0(E;\mathbf{r},\mathbf{r}')|&\simeq e^{\mu_{\text{max},M}\delta x}\sum_{n=-\infty}^{\infty}\exp[i\frac{2\pi (s_M\delta x-\delta y)}{L_y}n-\frac{4\pi^2 c_M\delta x}{L_y^2}n^2]\\
    &= e^{\mu_{\text{max},M}\delta x}\sum_{n=-\infty}^{\infty}\exp[i2\pi\alpha(s_M-g)n-\frac{4\pi^2 c_M\alpha^2}{L_x}n^2]\\
    &=e^{\mu_{\text{max},M}\delta x}\vartheta_3\left(\pi\alpha(s_M-g),e^{-\frac{4\pi^2 c_M \alpha^2}{L_x}}\right).
\end{aligned}
\end{equation}
Similarly, for $G_{0}(E;\mathbf{r}',\mathbf{r})$, since we are considering the thermodynamic limit, $|\delta x|$ is sufficiently large such that $M-1-\delta x< 0$. In this case, $\beta _{x} =0$ becomes a pole of order $|M-1-\delta x|$ and $\beta _{x} =\infty $ is no longer a pole. Therefore, we consider the contribution of the poles outside GBZ on the complex plane and we get
\begin{equation}
    G_{0}(E;\mathbf{r}',\mathbf{r})=\sum _{j=M+1}^{M+N}\int _{-\pi }^{\pi }\mathrm{d} k_{y}\frac{1}{[ \beta _{j}( k_{y})]^{|M-1-\delta x|}}\frac{-e^{-ik_{y} \delta y}}{( \beta _{j}( k_{y}) -\beta _{M+N}( k_{y})) \cdots ( \beta _{j}( k_{y}) -\beta _{1}( k_{y}))}.
\end{equation}
The dominant contribution comes from the branch with the smallest magnitude, i.e. the $\beta _{M+1}$ branch. We also omit the $( \beta _{j}( k_{y}) -\beta _{M+N}( k_{y})) \cdots ( \beta _{j}( k_{y}) -\beta _{1}( k_{y}))$ terms (since it does not scale with system size) and get an estimation:
\begin{equation}
    G_0(E;\mathbf{r}',\mathbf{r})\simeq\int_{-\pi}^{\pi}\dd k_y\left(\frac{1}{\beta_{M+1}(k_y)}\right)^{|\delta x+1-M|}e^{-i k_y\delta y}.
\end{equation}
We can also expand $\beta_{M+1}(k_y)$ as
\begin{equation}
    \beta_{M+1}(k_y)=\exp{\mu_{\text{min},M+1}+is_{M+1}(k_y-k_m)+c_{M+1}(k_y-k_m)^2}.
\end{equation}
Taking this expansion into $G_0(E;\mathbf{r}',\mathbf{r})$ and finally we get
\begin{equation}
    |G_0(E;\mathbf{r}',\mathbf{r})|\simeq e^{-\mu_{\text{min},M+1}\delta x}\vartheta_3\left(\pi\alpha(s_{M+1}-g),e^{-\frac{4\pi^2 c_{M+1}\alpha^2}{L_x}}\right)
\end{equation}
Therefore, the amplitude of the forth-back propagator is given by
\begin{equation}\label{eq:generic_approx}
    |G_0(E;\mathbf{r},\mathbf{r}')G_0(E;\mathbf{r}',\mathbf{r})|=e^{(\mu_{\text{max},M}-\mu_{\text{min},M+1})\delta x}\vartheta_3\left(\pi\alpha(s_{M+1}-g),e^{-\frac{4\pi^2 c_{M+1}\alpha^2}{L_x}}\right)\vartheta_3\left(\pi\alpha(s_M-g),e^{-\frac{4\pi^2 c_M \alpha^2}{L_x}}\right).
\end{equation}
This result indicates that for a generic model with long-range hoppings, among all of the solutions branches $\{\beta_i:i=1,\cdots, M+N\}$, only the two closest branches to the GBZ, i.e. branch $\beta_M$ and $\beta_{M+1}$ determines the sensitivity in the thermodynamic limit.
Moreover, the result can also be approximately represented by a Jacobi $\vartheta_3$ function with geometric parameters $(\alpha,g)$ as we showed in the Method I sections.

\subsection{II.2~Maximally sensitive region for reciprocal models}
For reciprocal models, due to the reciprocity of the Hamiltonian $\hat{H}_0=\hat{H}_0^T$, we have $|G_0(E;\mathbf{r},\mathbf{r}')|=|G_0(E;\mathbf{r}',\mathbf{r})|$.
This is also reflected in the symmetry between the $\beta_{M}$ and $\beta_{M+1}$ solutions where $s_M=s_{M+1}$, $c_M=c_{M+1}$, and $\mu_{\text{max},M}=-\mu_{\text{min},M+1}$.
Therefore, we only need to consider the divergence of the forward propagation $|G_0(E;\mathbf{r},\mathbf{r}')|$ and the backward propagation is automatically taken care of.
Then applying the same analysis for the $\beta_M$, we can reproduce the phase diagram in the $(\alpha,g)$ plane with essentially the same physics as the hopping-range-1 reciprocal model with only $\beta_1$ and $\beta_2$ solutions.

For reciprocal models, we can always fine-tune the positions of the bi-impurities such that $\delta y=s_M\delta x$ to activate the sensitivity maximally.
For a given energy $E$ outside the spectral region of $\hat{H}_0$, if the sensitive condition is satisfied, then we call this energy $E$ belongs to the spectral sensitive region.
We define the collection of these energies on the complex plane as the maximally sensitive region $\sigma_{\text{sens}}$.
\begin{equation}
    \sigma_{\text{sens}}:=\left\{E\in\mathbb{C}:\,\rho(\hat{G}_0(E)\hat{V}_{\delta y=s\delta x})>1\right\}.
\end{equation}
We also define the spectral area of the system under PBC as $\sigma_{\text{PBC}}$:
\begin{equation}
    \sigma_{\text{PBC}}:=\left\{E\in\mathbb{C}:\,E=\mathcal{H}_0(k_x,k_y)\text{ for some }(k_x,k_y)\in\text{BZ}\right\}.
\end{equation}
Now we claim that for reciprocal systems, the maximally sensitive region is exactly the spectral region $\sigma_{\text{sens}}=\sigma_{\text{PBC}}$.
Actually, when $\delta y=s_M\delta x$, we have $|G_0(E;\mathbf{r},\mathbf{r}')|\sim e^{\mu_{\text{max},M}\delta x}$. Therefore, $E\in\sigma_{\text{sens}}$ is essentially requiring that $\mu_{\text{max},M}>0$.
Since $\mu_M(k_y)=\log|\beta_M(k_y)|$ is a continuous curve on $k_y\in[-\pi,\pi]$, if $\mu_{\text{max}, M}>0$, then the curve must have zeros, i.e. $\exists k_y^*$ such that $|\beta_M(k_y^*)|=1$. 
So, in other words, there must exists $k_x^*,k_y^*\in[0,2\pi]$ such that $\det|E\mathbb{I}-\mathcal{H}_0(k_x^*,k_y^*)|=0$.
Hence, $E\in\sigma_{\text{PBC}}$. This statement is also true conversely, so $\sigma_{\text{sens}}=\sigma_{\text{PBC}}$.

\begin{figure}[h]
    \centering
    \includegraphics[width=0.8\linewidth]{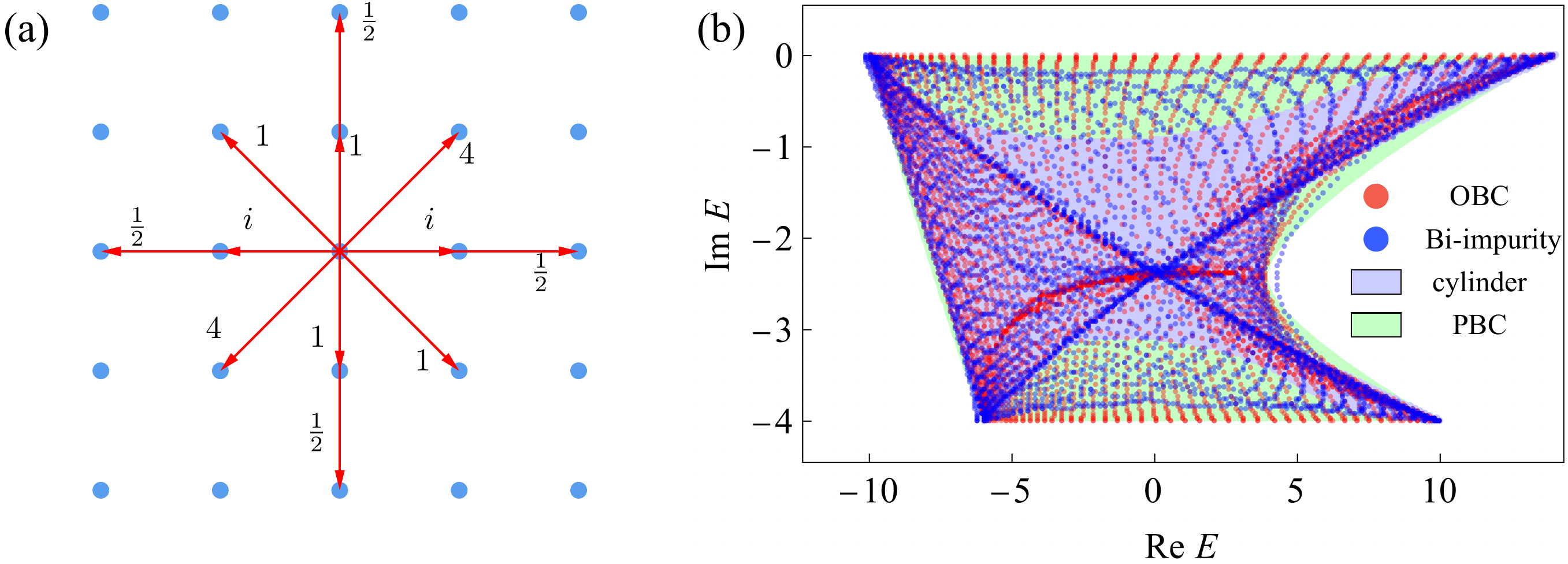}
    \caption{\textbf{Spectrum of a reciprocal model with next-to-nearest-neighbor hopping under OBC, PBC, cylinder, and bi-impurity cylinder geometry.}
    (a) sketch showing the hopping strength of the model.
    (b) The OBC (red dots), PBC (light green region), cylinder (light blue region), and bi-impurity cylinder (blue dots) spectrum for a rectangle of size $(L_x, L_y)=(100,50)$.
    }
    \label{fig:sf3}
\end{figure}

Here, we exemplify this with a tight-binding non-Hermitian model with long-range hoppings. The hopping parameters are shown in Fig.~\ref{fig:sf3} (a).
We calculate the spectrum of this model on a system with sufficiently large $\alpha$ so that the sensitive region is maximally excited.
As is depicted in Fig.~\ref{fig:sf3} (b), the coverage of the OBC and bi-impurity cylinder spectrum is almost identical to the PBC spectral coverage (minor offset due to finite-size effect).

\section{III.~Non-reciprocal models}
In non-reciprocal systems, an additional spectral sensitivity arises when cutting periodic to open boundary conditions~\cite{Budich2020PRL,budich2023}.
For instance, in the 1D Hatano-Nelson model, tuning the boundary link of PBC from $0^+$ to $0$ disrupts the spectral flow induced by non-reciprocal hopping, leading to a drastic contraction of the spectrum from a loop to an arc.
In 2D systems, this effect causes the OBC spectrum to shrink, potentially failing to cover certain regions of the spectrum under cylindrical geometry. 
in 2D non-reciprocal systems, this spectral sensitivity coexists with the bi-impurity sensitivity that we explore in this work.

In the main text, we relate the bi-impurity cylinder spectrum to the OBC spectrum through the following reasoning:
\begin{enumerate}
    \item Introduce an impurity line aligned with the bi-impurity direction;
    \item Vary the impurity potential strength $W$ from $0^+$ to $+\infty$, which effectively cuts the boundary links, leading to OBC (since the spatial profile of eigenstates with finite energy must vanish at the impurity line).
\end{enumerate}
With the competing spectral sensitivity in non-reciprocal systems,  step 2 is invalidated as this competing sensitivity effect becomes prominent at $W = +\infty$, preventing us from adiabatically connecting the OBC spectrum to the bi-impurity or impurity-line spectrum.

\begin{figure}[h]
    \centering
    \includegraphics[width=0.9\linewidth]{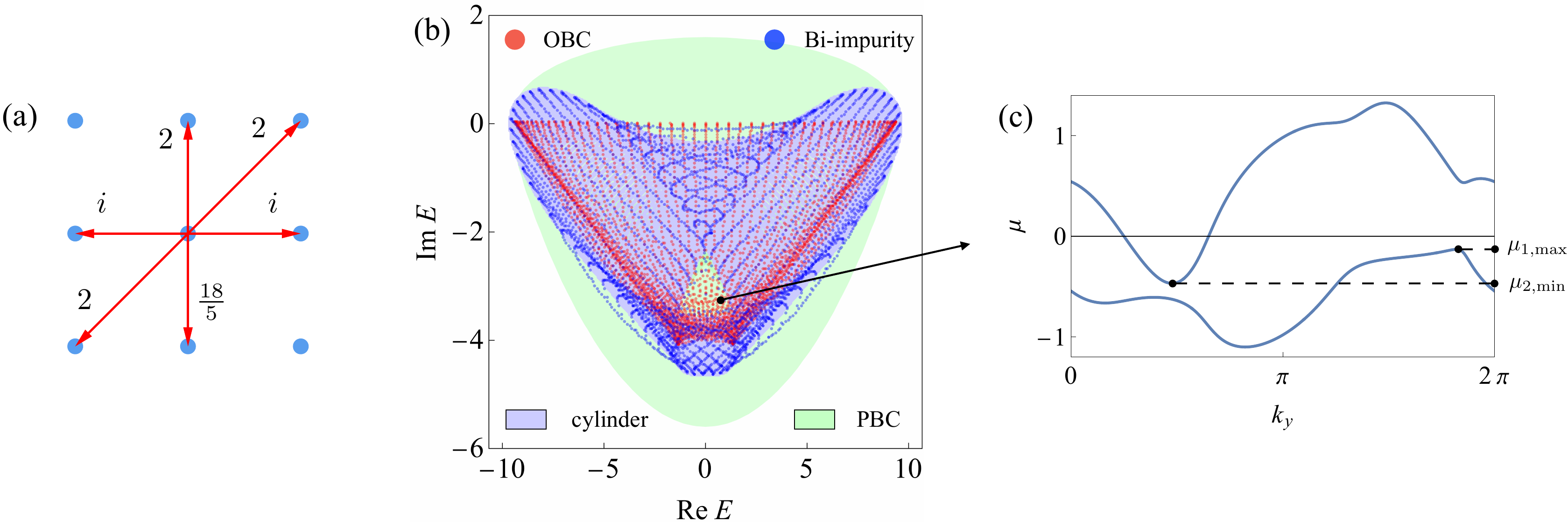}
    \caption{\textbf{Spectral sensitivity of a non-reciprocal model.}
    (a) Illustration of the hopping strength.
    (b) Spectrum under PBC (light green region), OBC (red dots), cylinder geometry (light blue region), and bi-impurity cylinder (blue dots). The system size is $(L_x,L_y)=(70,50)$.
    (c) The $\beta(k_y)$ solution curves corresponding to an energy point (marked as a black dot in (b)) in the "hole" of the cylinder spectrum.
    }
    \label{fig:sf4}
\end{figure}

However, the criterion relating to the divergence of the forth-back propagator Eq.~\eqref{eq:generic_approx} still applies to the energy beyond the cylinder spectrum $E\notin\sigma_{\text{cyl}}$, so that the two solution branches are still gapped.
When reciprocity is absent, the divergence of $|G_0(E;\mathbf{r},\mathbf{r}')|$ and $|G_0(E;\mathbf{r}',\mathbf{r})|$ has to be considered respectively.
Since $s_M\neq s_{M+1}$, the two impurities can no longer be perfectly aligned to activate the sensitive as in the reciprocal case maximally.
Therefore, distinct $s_M$ and $s_{M+1}$ will result in a more complicated phase diagram, which can be obtained by performing asymptotic expansion of the Eq.~\eqref{eq:generic_approx}.
Nevertheless, with sufficiently large $\alpha$, the spectrum will still be sensitive to the bi-impurity.

Again, we provide a non-reciprocal model as an example to illustrate the points.
The model and its spectrum are depicted in Fig.~\ref{fig:sf4}.
One interesting feature of this model is that the cylinder spectrum (light blue region in Fig.~\ref{fig:sf4} (b)) has a hole.
When putting bi-impurity to the cylinder (with sufficiently large $\alpha$) and re-calculate the spectrum (blue dots in  Fig.~\ref{fig:sf4} (b)), some spectral dots enter this hole, indicating that this region might be sensitive.
Applying our criterion and plotting the $\beta_M$ and $\beta_{M+1}$ branches in Fig.~\ref{fig:sf4} (c), we find that they satisfy the condition $\mu_{1,\text{max}}>\mu_{2,\text{min}}$.
With sufficiently large $\alpha$, that would imply the divergence of the forth-back propagator Eq.~\eqref{eq:generic_approx}.
Now, we calculate the spectrum under OBC (red dots in  Fig.~\ref{fig:sf4} (b)).
Indeed, the competing sensitivity effect renders partial shrinking of the OBC spectrum so that not all cylinder spectral area is covered by the OBC spectral region.
The coverage of the OBC spectrum is also significantly different from the bi-impurity cylinder spectrum.
To obtain the OBC spectrum analytically, we can use the boundary matrix method provided in Ref.~\cite{KaiASE2024}.

\subsection{III.1~The absence of sensitivity in purely exponential skin effect}
When the spectrum of a non-Hermitian system forms an arc, the skin modes must exhibit purely exponential decay from the boundary~\cite{KaiASE2024}.
Meanwhile, according to the spectral momentum theorem, the density of states $D(E)$ for this type of system must be insensitive to the boundary condition~\cite{Nan2024}.
Therefore, we conclude that such systems have no bi-impurity spectral sensitivity.

\begin{figure}
    \centering
    \includegraphics[width=\linewidth]{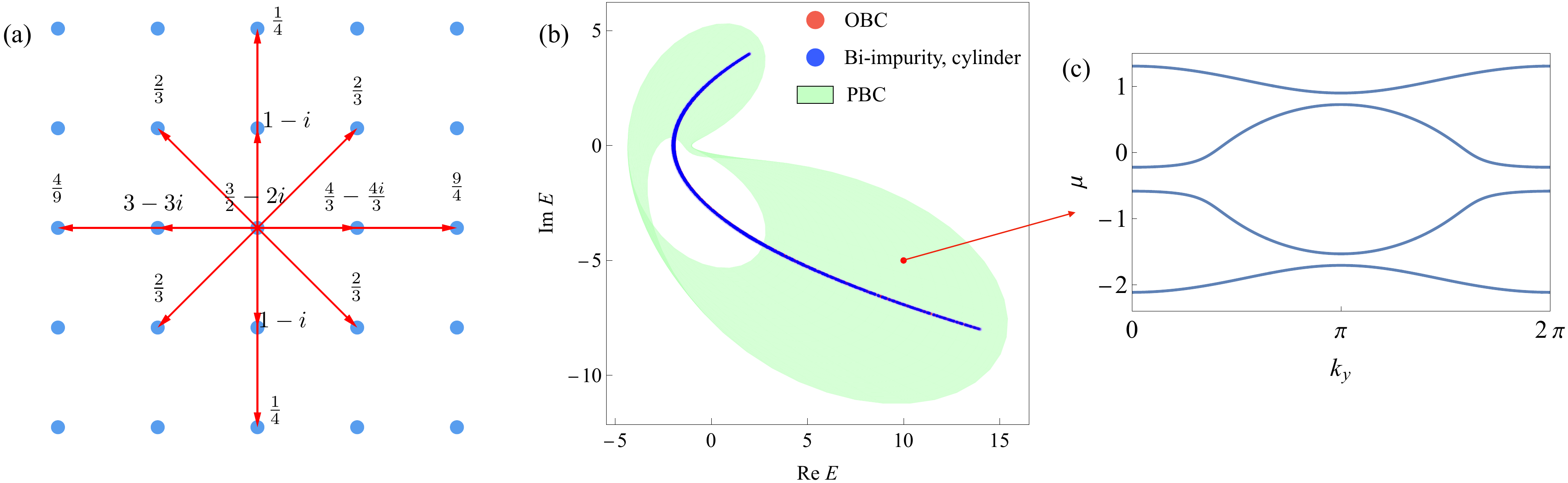}
    \caption{\textbf{Absence of sensitivity for non-reciprocal models with arc spectrum}.
    (a) Illustration of the hopping strength.
    (b) Spectrum under PBC (light green region), cylinder and bi-impurity cylinder (blue dots forming an arc), and OBC (red dots, invisible because they are covered by the blue dots). The system size is $(L_x,L_y)=(50,50)$ rectangle.
    (c) The $\beta(k_y)$ solutions corresponding to the red energy point $E=10-5i$ in (b). The $\beta_2$ and $\beta_3$ solutions are gapped and do not satisfy the sensitivity condition.
    }
    \label{fig:sf5}
\end{figure}

As an example, we present a general non-reciprocal model given by Fig.~\ref{fig:sf5}, whose OBC spectrum forms an arc.
In this case, the cylinder spectrum and bi-impurity cylinder spectrum both coincide with the OBC spectrum, confirming the absence of spectral sensitivity. (Fig.~\ref{fig:sf5} (b))
Correspondingly, we can also verify the absence of non-local diverging response.
For a random complex energy outside of the arc spectrum, the $\beta_{M}$ and $\beta_{M+1}$ solution curves are fully gapped and do not satisfy the sensitivity condition $\mu_{M,\text{max}}>\mu_{M+1,\text{min}}$. (Fig.~\ref{fig:sf5} (c))

\section{IV.~Bi-impurity induced algebraic skin modes}
The eigenstates of $\hat{H}_0$ in the main text are exponentially localized near the edge under cylinder geometry since the translational symmetry reduces the system to 1D NHSE.
Here, we show that by introducing the bi-impurity, the localization of the induced eigenstates beyond the cylinder spectrum becomes power-law instead of exponential.

\begin{figure}[h]
    \centering    
    \includegraphics[width=0.7\linewidth]{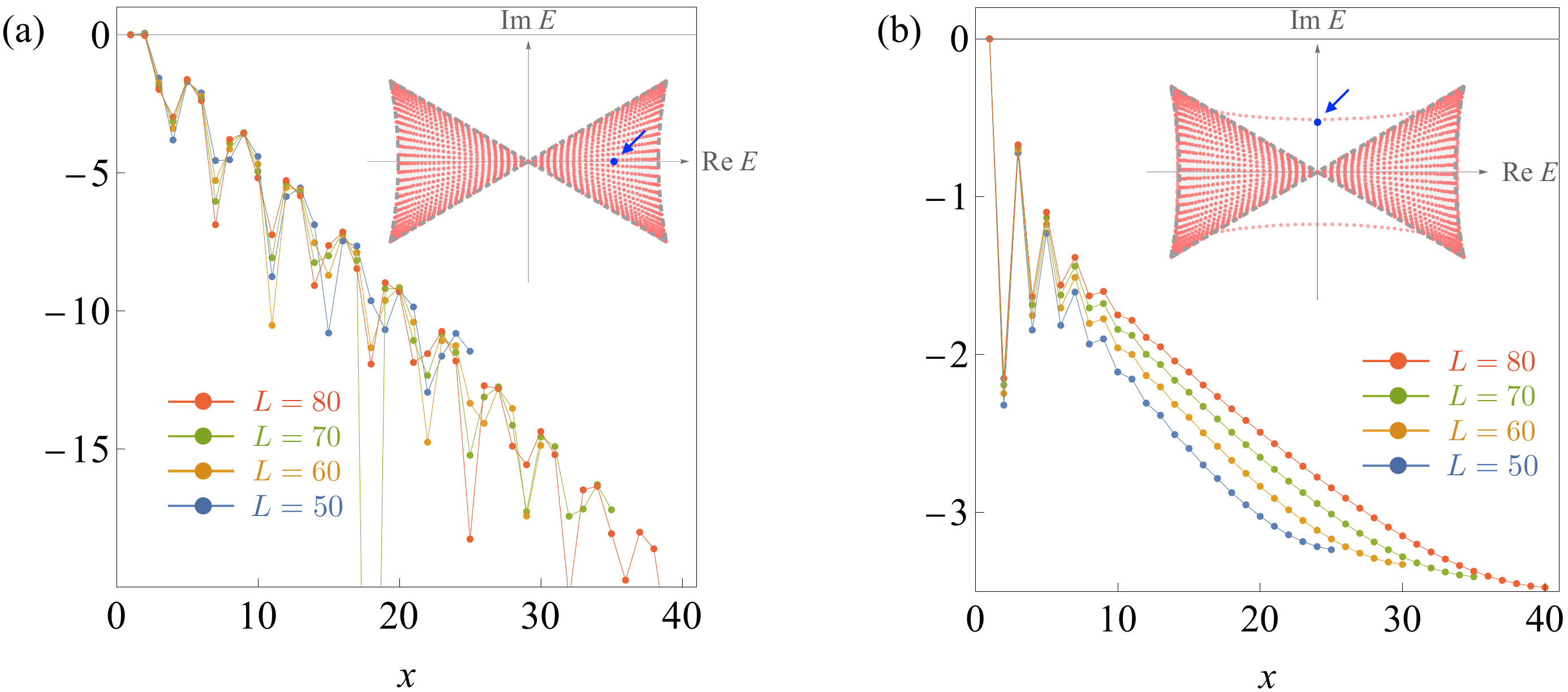}
    \caption{\textbf{Finite-size scaling of the algebraic skin modes induced by bi-impurity.}
    The dotted line represents the decay of the wavefunction layer density $P(x)$ away from the left boundary to the system's size.
    The system size is $L_x=L_y=L$ with $L=40,50,60,70$.
    (a) shows the decay of the $\log P(x)$ for the eigenenergy $E\simeq 5-2i$ of $H_0$ on the cylinder of different sizes. The blue point in the spectrum marks the chosen energy.
    The average decay rate is constant and does not change with system size, indicating that $P(x)$ decays exponentially.
    (b) shows the decay of the $\log P(x)$ for the eigenenergy $E=-0.8i$ of $H=H_0+V$ where $V$ is the bi-impurity potential. Now, with different $L$, the tail of the wavefunction shifts, indicating the power-law decay.
    }
    \label{fig:sf6}
\end{figure}

To study the wavefunction decay, we define the wavefunction layer density by summing up the amplitude square of each $x=\text{const.}$ layer and denote it as $P(x)$, given by
$
    P_{E,L}(x)=\sum_{y=0}^{L}|\psi_{E}(x,y)|^2
$
where $E$ is the chosen complex energy, $L$ is the system size, and $\psi_E$ is the eigenstate with energy $E$.
We properly normalize $P(x)$ by $\tilde{P}(x)=P(x)/P(0)$ such that for different system sizes $L$, the layer densities agree on the first layer.
If the wavefunction follows exponential decay $P(x)\sim e^{-x/\lambda}$, then the localization tail $\lambda$ does not depend on the system size.
As shown in Fig.~\ref{fig:sf6} (a), when we plot the $\log|P(x)|$, the slope $1/\lambda$ agrees with varying system sizes, confirming the exponential localization.
However, for the power-law decay, the characteristic length of the localization tail does scale with system size, due to the normalization and the divergence of the integration $\int x^{-a}\dd x$.
As shown in Fig.~\ref{fig:sf6} (b), the localization length of the tail increases as we enlarge the system size, signaling the algebraic decay.

\end{document}